\documentclass[
  journal=largetwo,
  manuscript=research-article,
  year=2023,
  volume=37,
]{cup-journal}

\usepackage{amsmath}
\usepackage[nopatch]{microtype}
\usepackage{booktabs}
\usepackage{graphicx}
\usepackage{hyperref}
\hypersetup{
    colorlinks=true,
    linkcolor=blue,
    filecolor=magenta,      
    urlcolor=cyan,
    citecolor=red
    }
\usepackage{txfonts}
\usepackage{multirow}
\usepackage{amsmath,amssymb,amsfonts}
\usepackage{manyfoot}
\usepackage{url}
\usepackage{longtable}
\DeclareUnicodeCharacter{0308}{\textendash}

\title{VaTEST III: Validation of 8 Potential Super-Earths from TESS Data}

\author{Priyashkumar Mistry} 
\affiliation{Department of Physics, Sardar Vallabhbhai National Institute of Technology, Surat-395007, Gujarat, India}
\alsoaffiliation{School of Physics, Faculty of Science, University of New South Wales, Kensington, Sydney NSW 2052, Australia}
\email[Priyashkumar Mistry]{priyashmistry10@gmail.com}

\author{Aniket Prasad} 
\affiliation{Department of Earth and Space Science, Indian Institute of Space Science and Technology, Thiruvananthapuram, India}

\author{Mousam Maity} 
\affiliation{Department of Physics, Presidency University, Kolkata-700073, West Bengal, India}

\author{Kamlesh Pathak}
\affiliation{Department of Physics, Sardar Vallabhbhai National Institute of Technology, Surat-395007, Gujarat, India}

\author{Sarvesh Gharat} 
\affiliation{Centre for Machine Intelligence and Data Science, Indian Institute of Technology Bombay}

\author{Georgios Lekkas} 
\affiliation{Department of Computer Science, International Hellenic University, 65404, Kavala, Greece}

\author{Surendra Bhattarai} 
\affiliation{Department of Physics, Indian Institute of Science Education and Research Kolkata, Mohanpur-741246, West Bengal, India}

\author{Dhruv Kumar} 
\affiliation{Department of Physics, National Institute of Technology Agartala 799046, Tripura, India}

\author{Jack J. Lissauer}
\affiliation{NASA Ames Research Center, Moffett Field, CA 94035, USA}

\author{Joseph D. Twicken}
\affiliation{NASA Ames Research Center, Moffett Field, CA 94035, USA}
\alsoaffiliation{SETI Institute, Mountain View, CA 94043, USA}

\author{Abderahmane Soubkiou}
\affiliation{Oukaimeden Observatory, High Energy Physics and Astrophysics Laboratory, Cadi Ayyad University, Marrakech, Morocco}

\author{Francisco J. Pozuelos}
\affiliation{Astrobiology Research Unit, Universit\'e de Li\`ege, 19C All\'ee du 6 Ao\^ut, 4000 Li\`ege, Belgium}
\alsoaffiliation{Instituto de Astrof\'isica de Andaluc\'ia (IAA-CSIC), Glorieta de la Astronom\'ia s/n, 18008 Granada, Spain}

\author{Jon Jenkins}
\affiliation{NASA Ames Research Center, Moffett Field, CA 94035, USA}

\author{Keith Horne}
\affiliation{SUPA Physics and Astronomy, University of St. Andrews, Fife, KY16 9SS Scotland, UK}

\author{Steven Giacalone}
\affiliation{Department of Astronomy, University of California Berkeley, Berkeley, CA 94720, USA}

\author{Khalid Barkaoui}
\affiliation{Astrobiology Research Unit, Universit\'e de Li\`ege, 19C All\'ee du 6 Ao\^ut, 4000 Li\`ege, Belgium}
\alsoaffiliation{Department of Earth, Atmospheric and Planetary Science, Massachusetts Institute of Technology, 77 Massachusetts Avenue, Cambridge, MA 02139, USA}                                 
\alsoaffiliation{Instituto  de  Astrofísica  de  Canarias  (IAC),  Calle  Vía  Láctea  s/n, 38200, La Laguna, Tenerife, Spain}

\author{Mathilde Timmermans}
\affiliation{Astrobiology Research Unit, Universit\'e de Li\`ege, 19C All\'ee du 6 Ao\^ut, 4000 Li\`ege, Belgium}

\author{Cristilyn N. Watkins}
\affiliation{Center for Astrophysics \textbar \ Harvard \& Smithsonian, 60 Garden Street, Cambridge, MA 02138, USA}

\author{Ramotholo Sefako}
\affiliation{South African Astronomical Observatory, P.O. Box 9, Observatory, Cape Town 7935, South Africa}

\author{Karen A.\ Collins}
\affiliation{Center for Astrophysics \textbar \ Harvard \& Smithsonian, 60 Garden Street, Cambridge, MA 02138, USA}

\author{David R. Ciardi} 
\affiliation{NASA Exoplanet Science Institute, IPAC, California Institute of Technology, Pasadena, CA 91125 USA}

\author{Catherine A. Clark} 
\affiliation{Jet Propulsion Laboratory, California Institute of Technology, Pasadena, CA 91109 USA}
\alsoaffiliation{NASA Exoplanet Science Institute, IPAC, California Institute of Technology, Pasadena, CA 91125 USA}

\author{Boris S. Safonov}
\affiliation{Sternberg Astronomical Institute Lomonosov Moscow State University}

\author{Avi  Shporer}
\affiliation{Department of Physics and Kavli Institute for Astrophysics and Space Research, Massachusetts Institute of Technology, Cambridge, MA 02139, USA}

\author{Joshua E. Schlieder}
\affiliation{NASA Goddard Space Flight Center, 8800 Greenbelt Road, Greenbelt, MD 20771, USA}

\author{Zouhair Benkhaldoun}
\affiliation{Oukaimeden Observatory, High Energy Physics and Astrophysics Laboratory, Cadi Ayyad University, Marrakech, Morocco}

\author{Chris Stockdale}
\affiliation{Hazelwood Observatory, Australia}

\author{Carl Ziegler}
\affiliation{Department of Physics, Engineering and Astronomy, Stephen F. Austin State University, 1936 North St, Nacogdoches, TX 75962, USA}

\author{Emily A. Gilbert}
\affiliation{Jet Propulsion Laboratory, California Institute of Technology, Pasadena, CA 91109 USA}

\author{Emmanu\"el Jehin}
\affiliation{Space Sciences, Technologies and Astrophysics Research (STAR) Institute, Universit\'e de Li\`ege, All\'ee du 6 Ao\^ut 19C, B-4000 Li\`ege, Belgium}

\author{Felipe Murgas}
\affiliation{Instituto de Astrof\'\i sica de Canarias (IAC), 38205 La Laguna, Tenerife, Spain}
\alsoaffiliation{Departamento de Astrof\'\i sica, Universidad de La Laguna (ULL), 38206, La Laguna, Tenerife, Spain}

\author{Ian J.\ M.\ Crossfield}
\affiliation{Department of Physics and Astronomy, University of  Kansas, Lawrence, KS, USA}

\author{Martin Paegert}
\affiliation{Center for Astrophysics \textbar \ Harvard \& Smithsonian, 60 Garden Street, Cambridge, MA 02138, USA}

\author{Michael B. Lund}
\affiliation{NASA Exoplanet Science Institute, IPAC, California Institute of Technology, Pasadena, CA 91125 USA}

\author{Norio Narita}
\affiliation{Komaba Institute for Science, The University of Tokyo, 3-8-1 Komaba, Meguro, Tokyo 153-8902, Japan}
\alsoaffiliation{Astrobiology Center, 2-21-1 Osawa, Mitaka, Tokyo 181-8588, Japan}
\alsoaffiliation{Instituto de Astrofisica de Canarias (IAC), 38205 La Laguna, Tenerife, Spain}

\author{Richard P.\ Schwarz}
\affiliation{Center for Astrophysics \textbar \ Harvard \& Smithsonian, 60 Garden Street, Cambridge, MA 02138, USA}

\author{Robert F. Goeke}
\affiliation{Department of Physics and Kavli Institute for Astrophysics and Space Research, Massachusetts Institute of Technology, Cambridge, MA 02139, USA}

\author{Sergio B. Fajardo-Acosta} 
\affiliation{Caltech/IPAC, 1200. E California Blvd., Pasadena, CA 91125}

\author{Steve~B.~Howell}
\affiliation{NASA Ames Research Center, Moffett Field, CA 94035, USA}

\author{Thiam-Guan Tan}
\affiliation{Perth Exoplanet Survey Telescope, Perth, Western Australia}

\author{Thomas Barclay} 
\affiliation{NASA Goddard Space Flight Center, 8800 Greenbelt Road, Greenbelt, MD 20771, USA}

\author{Yugo Kawai}
\affiliation{Department of Multi-Disciplinary Sciences, Graduate School of Arts and Sciences, The University of Tokyo, 3-8-1 Komaba, Meguro, Tokyo 153-8902, Japan}


\begin{document}

\begin{abstract}
NASA's all-sky survey mission, the Transiting Exoplanet Survey Satellite (TESS), is specifically engineered to detect exoplanets that transit bright stars. Thus far, TESS has successfully identified approximately 400 transiting exoplanets, in addition to roughly 6000 candidate exoplanets pending confirmation. In this study, we present the results of our ongoing project, the Validation of Transiting Exoplanets using Statistical Tools (VaTEST). Our dedicated effort is focused on the confirmation and characterization of new exoplanets through the application of statistical validation tools. Through a combination of ground-based telescope data, high-resolution imaging, and the utilization of the statistical validation tool known as \texttt{TRICERATOPS}, we have successfully discovered eight potential super-Earths. These planets bear the designations: TOI-238b (1.61$^{+0.09} _{-0.10}$ R$_\oplus$), TOI-771b (1.42$^{+0.11} _{-0.09}$ R$_\oplus$), TOI-871b (1.66$^{+0.11} _{-0.11}$ R$_\oplus$), TOI-1467b (1.83$^{+0.16} _{-0.15}$ R$_\oplus$), TOI-1739b (1.69$^{+0.10} _{-0.08}$ R$_\oplus$), TOI-2068b (1.82$^{+0.16} _{-0.15}$ R$_\oplus$), TOI-4559b (1.42$^{+0.13} _{-0.11}$ R$_\oplus$), and TOI-5799b (1.62$^{+0.19} _{-0.13}$ R$_\oplus$). Among all these planets, six of them fall within the region known as 'keystone planets,' which makes them particularly interesting for study. Based on the location of TOI-771b and TOI-4559b below the radius valley we characterized them as likely super-Earths, though radial velocity mass measurements for these planets will provide more details about their characterization. It is noteworthy that planets within the size range investigated herein are absent from our own solar system, making their study crucial for gaining insights into the evolutionary stages between Earth and Neptune.
\end{abstract}

\section{Introduction}\label{intro}
The science of exoplanet research has dramatically advanced in the last two decades since the discovery of HD 209458b \citep{2000ApJ...529L..45C,2000ApJ...529L..41H}, the first exoplanet to be detected using the transit method. The last two decades have seen an exponential growth in the number of exoplanets detected and yet many more remain to be validated with further observations. Our ability to explore the diverse exoplanet population has rapidly increased through scientific and technological advancements, improving telescope observational capabilities in both space-based and ground-based observations. The Transiting Exoplanet Survey Satellite \citep[TESS;][]{2015JATIS...1a4003R} has emerged as a pioneering mission for discovering new transiting planets in the vicinity of our solar system. Launched on 2018 April 18, TESS was set to observe the brightest stars near the Earth for transiting exoplanets over a two-year period. To date, it has detected about 400 exoplanets and yet about 6000 candidates remain unvalidated.

The launch of space missions like Kepler \citep{borucki2010kepler}, K2 (the second mission of the Kepler spacecraft) \citep{howell2014k2}, and TESS \citep{2015JATIS...1a4003R} have provided us with valuable data which resulted in the discovery of a huge number of exoplanets. However, one of the most common problems that transiting exoplanet searches face is the detection of so-called False Positive events. False positives or false detection occur when a transit signal from the target is caused by something other than a true exoplanet transit, such as a background source or eclipsing binaries.  False detection in transit space missions can be mitigated by statistically validating transit candidates. However, it's important to note that even with rigorous validation methods the possibility of some candidates turning out to be false positives cannot be entirely eliminated. Significant efforts have been made to develop efficient statistical validation tools that can be used for a variety of space missions. And many planets have been validated using such tools to date. The list of such tools contains \texttt{BLENDER} \citep{2005ApJ...619..558T}, \texttt{PASTIS} \citep{2014MNRAS.441..983D}, \texttt{VESPA} \citep{2015ascl.soft03011M}, and \texttt{TRICERATOPS} \citep{2020ascl.soft02004G,2021AJ....161...24G}. \texttt{VESPA} and \texttt{TRICERATOPS} are the most commonly used tools in the Kepler and TESS era respectively \citep{2016ApJ...822...86M,2021AJ....161...24G,2022AJ....163..244C,2022AJ....163...99G,2023MNRAS.521.1066M,2023AJ....166....9M}. As per \citet{2023RNAAS...7..107M}, \texttt{VESPA} is now retired as it is no longer maintained and has not been updated to account for the modern astronomy data landscape. They recommended using the actively maintained \texttt{TRICERATOPS} package for statistical validation. So in this paper, we made use  of \texttt{TRICERATOPS} to validate the planetary nature of the transiting signal.

This research is a part of the Validation of Transiting Exoplanets using Statistical Tools (VaTEST)\footnote{\url{https://sites.google.com/view/project-vatest/home}} project, which aims to validate new extra-solar planetary systems using various statistical as well as machine learning based tools. We also characterize the validated exoplanets for their further atmospheric study either using space or ground-based observations. TOI-181b \citep{2023MNRAS.521.1066M} was the very first exoplanet promoted from a planetary candidate to a validated planet by this project. This was a sub-Saturn (smaller than Saturn) with the largest H/He envelope among all the known sub-Saturns and was massive enough to survive the photoevaporation. Our second paper was about the validation of 11 new exoplanets orbiting K spectral type stars \citep{2023AJ....166....9M}. In the mentioned study, we identified several systems conducive to atmospheric characterization through different spectroscopic techniques. These include TOI-2194b for transmission spectroscopy, TOI-3082b and TOI-5704b for emission spectroscopy, and TOI-672b, TOI-1694b, and TOI-2443b suitable for both transmission and emission spectroscopy.
 
In this paper, we aim to study the potential super-Earths (radii between 1.25 and 2 R$_\oplus$). It's worth noting that planets falling within this range are not present in our own solar system. The study of such planets is crucial for gaining insights into the evolutionary stages that bridge the gap between Earth and Neptune. We use statistical validation tools along with ground-based transit follow-up observations and high resolution imaging to validate the existence of exoplanets. Four of our validated planets i.e., TOI-238b, TOI-871b, TOI-1739b, and TOI-5799b are part of a region called Radius Valley. Radius valley is a region between 1.5-1.8~$R_{\oplus}$ in the exoplanet population \citep{2017AJ....154..109F, 2013ApJ...775..105O}. Although we do not present mass measurements in this paper, this could be done with high-precision radial velocity observation, as discussed in Section \ref{sec:RV Follow-up}. 

Our paper is organized as follows: In section \ref{sec:TESS Data and Candidate Selection}, we present the utilization of TESS data, candidate selection, and stellar parameters. We present our ground-based follow-up observations in section \ref{sec:Validation Techniques} and validation method in section \ref{sec:triceratops}, then present the planetary and orbital parameters of validated systems in section \ref{sec:Characteristics of the Validated Systems}. In section \ref{sec:discussion}, we illustrate various interesting features of our validated systems. We conclude our paper in section \ref{sec:conclusion}.

\section{TESS Data and Candidate Selection}\label{sec:TESS Data and Candidate Selection}
We use 2-minute cadence photometric data for our analysis. These data were collected using TESS and processed by the TESS Science Processing Operations Center \citep[TESS-SPOC;][]{2016SPIE.9913E..3EJ} pipeline and made available in the form of target pixel files (TPF) and light curve files including Presearch Data Conditioning Simple Aperture Photometry \citep[PDCSAP;][]{2012PASP..124..985S,2014PASP..126..100S,2012PASP..124.1000S} that is cleaned from instrumental systematics. The SPOC Transiting Planet Search \citep[TPS;][]{2002ApJ...575..493J,2010SPIE.7740E..0DJ,2020ksci.rept....9J} module employs an adaptive, noise-compensating matched filter, and was responsible for the recovery of the transit signals for each candidate validated here. Transit search Threshold Crossing Events (TCEs) were fitted with an initial limb-darkened transit model \citep{2019PASP..131b4506L}, and a suite of diagnostic tests were conducted to help assess the planetary nature of the signals \citep{2018PASP..130f4502T}. The TESS Science office reviewed the vetting information and promoted the TCEs to TESS Object of Interest (TOI) planet candidate status \citep{2021ApJS..254...39G} based on clean data validation reports.

We used the following criteria for our sample selection:
\begin{enumerate}
    \item As we were looking for potential super-Earths, we selected 317 TOI planet candidates with reported a radius between 1.25 and 2.00 R$_\oplus$ from the Exoplanet Follow-up Observing Program website ({\tt ExoFOP})\footnote{\url{https://exofop.ipac.caltech.edu/tess/}, accessed on 19th June, 2023}.
    \item Then we removed 111 candidates marked with eclipsing binary or false positive on the {\tt ExoFOP} website.
    \item We also discarded the candidates for which high-resolution imaging has detected any nearby companion.
    \item \texttt{Juliet} modeling is performed on the rest of the candidates to identify possible eclipsing binaries based on the shape (V-shaped) and characteristics of modeled transit light curves.
    \item Further we check the dispositions of 133 selected candidates provided by TESS Follow-up Observation Program Sub Group 1 \citep[TFOP SG1;][]{collins:2019}. We select eight candidates with disposition cleared planetary candidate (CPC), verified planet candidate (VPC) or verified planet candidate plus (VPC+). More details on these dispositions are mentioned in Section \ref{sec:Ground_Based_Phtotmetry}.
\end{enumerate}
Provenance data for these eight candidates are given in Table \ref{tab:provenance}. In the following sections, we discuss their validation techniques, planetary parameters, and important features of these planets.

\renewcommand{\arraystretch}{1.2}
\begin{table*}
    \centering
    \resizebox{\textwidth}{!}{
    \begin{tabular}{c c c c c}
    \hline
    & TOI-238.01 & TOI-771.01 & TOI-871.01 & TOI-1467.01 \\ 
    \hline
    \hline
    TIC ID & 09006668 & 277634430 & 219344917 & 240968774 \\
    Sectors Observed & 2, 29 & 10-12, 37, 38, 64 & 4-6, 31, 32 & 17, 18, 58\\
    SPOC Detection & Sector 2 (2018-10-04) & Sector 10 (2019-05-23) & Sectors 4-6 (2019-04-18) & Sector 17 (2019-11-15)\\
    TOI Alert & 2018-11-29 & 2019-06-05 & 2019-07-12 & 2019-12-05\\
    Identified Period (days) & 1.27 & 2.33 & 28.69\footnotemark[1] & 5.97\\ 
    & & & & \\
    \hline
    & TOI-1739.01 & TOI-2068.01 & TOI-4559.01 & TOI-5799.01 \\
    \hline
    \hline
    TIC ID & 159418353 & 417931300 & 271169413 & 328081248 \\
    Sectors Observed & 14, 19-21, 25, 26, 40, 41, 47 & 14, 15, 21, 22, 41, 48 & 11, 38 & 54\\
    SPOC Detection & Sectors 14 \& 19 (2020-01-24) & Sector 22 (2020-05-05) & Sectors 11 \& 38 (2021-07-22) & Sector 54 (2022-08-18)\\
    TOI Alert & 2020-02-27 & 2020-07-15 & 2021-10-28 & 2022-09-22\\
    Identified Period (days) & 8.30 & 7.77 & 3.96 & 4.16\\
    \hline
    \end{tabular}}
    \footnotemark[1]{The orbital period was subsequently refined to 14.36 d following the subsequent search of the light curve with all available data through sector 32 on 2021-05-27}
    \caption{Details of the observations and detection.}
    \label{tab:provenance}
\end{table*}
\renewcommand{\arraystretch}{1}

\subsection{Stellar Properties}
We determined the stellar parameters through a combination of spectral energy distribution (SED) analysis \citep{2016ApJ...831L...6S} and fitting MIST isochrones \citep{2016ApJ...823..102C,2016ApJS..222....8D} jointly with EXOFASTv2 \citep{2019arXiv190709480E}. This joint fitting approach provides precise measurements of the star's radius, mass, age, and surface gravity ($\log{g}$) \citep{2022arXiv220914301E}. To perform this analysis, we utilized TESS transit photometry data, broadband photometry, and the Gaia Data Release 3 (Gaia DR3) parallax information \citep{2023A&A...674A...1G}. The resulting stellar parameters are presented in Table \ref{tab:stellar}. Each entry in the Gaia database includes two valuable diagnostics for identifying unresolved binaries: the reduced unit weight error (RUWE) and the image parameter determination fraction of multiple peak (IPDfmp, \texttt{ipd\_frac\_multi\_peak} in Gaia terminology) \citep{2020MNRAS.496.1922B,2022MNRAS.513.5270P}. A RUWE value greater than 1.4 is generally considered indicative of deviations from a single-star astrometric solution, suggesting a possible unresolved binary. The IPDfmp parameter, which ranges from 0 to 100, serves as an even more potent diagnostic for identifying close companions compared to RUWE. However, it has received relatively little attention in the literature. Stars with an \texttt{ipd\_frac\_multi\_peak} value of 2 or lower are likely to be single stars. We have listed these values in Table \ref{tab:stellar}. It can be noted here that for TOI 1467 we reported a RUWE value that is greater than 1.4, but on the other hand, the \texttt{ipd\_frac\_multi\_peak} value is 0, which favors the case of single star solution. Also using high-resolution imaging techniques (as you will find in Section \ref{sec:Keck/NIRC2}), we detect no nearby companion to TOI 1467.

\renewcommand{\arraystretch}{1.5}
\begin{table*}
    \centering
    \resizebox{0.6\textwidth}{!}
    {
    \begin{tabular}{l c c c c}
    \hline
    Parameters & TOI 238      & TOI 771       & TOI 871       & TOI 1467      \\
               & TIC 09006668 & TIC 277634430 & TIC 219344917 & TIC 240968774 \\
    \hline
    \hline
    RA (J2000) & 23:16:55.46 & 10:56:27.33 & 04:58:57.03 & 01:16:27.51 \\
    
    Dec (J2000) & -18:36:23.9 & -72:59:06.6 & -50:37:38.6 & +49:13:59.3 \\
    
    Radius ($R_\odot$)     & $0.750^{+0.025}_{-0.026}$ & $0.242^{+0.012}_{-0.011}$       & $0.719\pm0.025$              & $0.472\pm0.018$ \\
    
    Mass ($M_\odot$)       & $0.788^{+0.047}_{-0.048}$ & $0.220^{+0.024}_{-0.023}$       & $0.758^{+0.046}_{-0.044}$    & $0.498^{+0.026}_{-0.025}$ \\
    
    Luminosity ($L_\odot$) & $0.3370\pm0.0210$           & $0.0055\pm0.0004$ & $0.2754^{+0.0076}_{-0.0077}$ & $0.0424^{+0.0024}_{-0.0028}$ \\
    
    Density (cgs)          & $2.64^{+0.25}_{-0.22}$    & $21.90^{+3.10}_{-2.70}$            & $2.87^{+0.26}_{-0.24}$       & $6.66^{+0.67}_{-0.59}$ \\
    
    Surface Gravity ($\log{g}$) & $4.584\pm0.029$           & $5.014^{+0.043}_{-0.044}$       & $4.604\pm0.027$              & $4.787\pm0.027$ \\
    
    Temperature (K)        & $5080\pm130$              & $3201^{+100}_{-95}$             & $4929^{+80}_{-75}$           & $3810^{+73}_{-78}$ \\
    
    Metallicity (dex)      & $-0.05^{+0.23}_{-0.26}$   & $0.12^{+0.19}_{-0.28}$          & $-0.07^{+0.23}_{-0.24}$      & $-0.10^{+0.27}_{-0.24}$ \\
    
    Age (Gyr)              & $6.0^{+5.2}_{-4.1}$       & $6.7^{+4.9}_{-4.8}$             & $5.9^{+5.0}_{-4.2}$          & $7.2^{+4.5}_{-4.9}$ \\
    
    Distance (pc)          & $80.53^{+0.33}_{-0.32}$   & $25.28\pm0.04$                  & $68.04\pm0.15$               & $37.44\pm0.06$ \\
    
    V Mag                  & $10.748 \pm 0.013$        & $14.888 \pm 0.080$              & $10.569 \pm 0.010$                   & $12.293 \pm 0.017$ \\
    
    TESS Mag               & $9.927 \pm 0.006$         & $12.087 \pm 0.007$              & $9.761 \pm 0.006$                   & $10.597 \pm 0.007$ \\

    J Mag & $9.214 \pm 0.024 $ & $10.507 \pm 0.023$ & $8.954 \pm 0.030$ & $9.380 \pm 0.018$ \\

    RUWE & 0.865 & 1.277 & 0.971 & 1.467 \\
    IPDFMP & 0 & 0 & 0 & 0 \\
    & & & & \\
    \hline
               & TOI 1739      & TOI 2068      & TOI 4559      & TOI 5799      \\
               & TIC 159418353 & TIC 417931300 & TIC 271169413 & TIC 328081248 \\
    \hline
    \hline
    RA (J2000) & 16:00:42.56 & 12:25:05.65 & 14:04:03.25 & 20:06:31.24 \\
    
    Dec (J2000) & +83:15:31.2 & +60:25:06.0 & -30:00:50.8 & +15:59:20.9 \\
    
    Radius ($R_\odot$)     & $0.751\pm0.024$           & $0.535\pm0.022$              & $0.374^{+0.017}_{-0.015}$    & $0.328\pm0.014$ \\
    Mass ($M_\odot$)       & $0.790^{+0.046}_{-0.045}$ & $0.559^{+0.027}_{-0.029}$    & $0.392^{+0.027}_{-0.026}$    & $0.337^{+0.027}_{-0.032}$ \\
    Luminosity ($L_\odot$) & $0.2990\pm0.0160$           & $0.0489^{+0.0023}_{-0.0024}$ & $0.0203^{+0.0013}_{-0.0016}$ & $0.0148^{+0.0013}_{-0.0012}$ \\
    Density (cgs)          & $2.63^{+0.24}_{-0.21}$    & $5.15^{+0.55}_{-0.50}$       & $10.50^{+1.10}_{-1.00}$         & $13.40^{+1.50}_{-1.40}$ \\
    Surface Gravity ($\log{g}$) & $4.584\pm0.028$           & $4.729^{+0.028}_{-0.029}$    & $4.884\pm0.030$              & $4.932^{+0.034}_{-0.036}$ \\
    Temperature (K)        & $4922^{+94}_{-91}$        & $3710^{+57}_{-58}$           & $3558^{+73}_{-83}$           & $3514^{+95}_{-94}$ \\
    Metallicity (dex)      & $0.07^{+0.24}_{-0.23}$    & $0.27^{+0.15}_{-0.20}$       & $-0.09^{+0.20}_{-0.22}$      & $-0.08^{+0.25}_{-0.34}$ \\
    Age (Gyr)              & $6.3^{+4.9}_{-4.3}$       & $7.0^{+4.6}_{-5.0}$          & $7.0^{+4.7}_{-4.6}$          & $7.0^{+4.7}_{-4.8}$ \\
    Distance (pc)          & $70.98\pm0.16$            & $52.93^{+0.09}_{-0.08}$      & $27.81\pm0.03$   & $27.81\pm0.03$ \\
    V Mag                  & $10.692 \pm 0.008$        & $13.007 \pm 0.009$           & $13.115 \pm 0.006$           & $13.290 \pm 0.077$ \\
    TESS Mag               & $9.812 \pm 0.006$         & $11.181 \pm 0.007$           & $10.914 \pm 0.007$           & $11.179 \pm 0.007$ \\
    J Mag & $8.982 \pm 0.021$ & $9.872 \pm 0.023$ & $9.455 \pm 0.024$ & $9.742 \pm 0.023$ \\
    
    RUWE & 0.997 & 1.161 & 1.357 & 1.184 \\
    IPDFMP & 0 & 2 & 0 & 0 \\
    \hline
    \end{tabular}
    }
    \caption{Stellar parameters derived using the \texttt{ExoFASTv2} tool \citep{2019arXiv190709480E}.}
    \label{tab:stellar}
\end{table*}
\renewcommand{\arraystretch}{1}

\section{Observations}\label{sec:Validation Techniques}

\subsection{Ground Based Photometry}\label{sec:Ground_Based_Phtotmetry}
The TESS pixel scale is approximately $\sim 21^{\prime \prime}$ pixel$^{-1}$, and photometric apertures extend to about 1$^{\prime}$, resulting in the blending of multiple stars within the TESS aperture. To eliminate the possibility of a nearby eclipsing binary (NEB) or a shallower nearby planet candidate (NPC) blend as the potential source of a TESS detection and to attempt to detect the signal on-target, we conducted observations of our target stars and the adjacent fields as part of the TFOP\footnote{\url{https://tess.mit.edu/followup}} SG1 \citep{collins:2019} initiative. In some instances, we also conducted observations in multiple spectral bands across the optical spectrum to check for wavelength-dependent transit signals, which could be indicative of a false positive planet candidate. To schedule our transit observations, we utilized the {\tt TESS Transit Finder}, a customized version of the {\tt Tapir} software package \citep{Jensen:2013}.

We have compiled all our light curve follow-up observations in Table \ref{tab:transitfollowup_VaTESTIII}, and the complete light curve dataset is accessible through the {\tt ExoFOP} website. In the subsequent sections, we describe each of the observatories employed to determine the final photometric outcomes. Additionally, a concise summary of each light curve result, along with an overall final photometric follow-up determination, is presented in Table \ref{tab:transitfollowup_VaTESTIII}. To convey our level of confidence regarding the on-target nature of a TESS detection, we employ three distinct light curve follow-up disposition codes, namely CPC, VPC, and VPC+, each indicating varying degrees of confidence, as elaborated below. 

The designation CPC signifies that we have effectively established that the TESS detection is associated with the target star rather than any other stars listed in the Gaia Data Release 3 \citep{2022yCat.1355....0G} and the TIC version 8 stars \citep{2019yCat.4038....0S}. Using ground-based photometry, we perform an extensive assessment of all stars located within a $2.5^{\prime}$ radius from the target star that exhibit sufficient brightness, assuming a 100\% eclipse in the TESS band, capable of generating the observed depth at mid-transit. To account for potential differences in magnitude between the TESS band and the follow-up band, as well as to accommodate magnitude errors specific to the TESS band, we include a buffer of 0.5 magnitudes fainter in the TESS band. In such cases, the transit depth is often too shallow to be reliably detected on the target star during ground-based follow-up observations. Consequently, we may intentionally saturate the target star on the detector to facilitate a comprehensive search of all nearby fainter stars. Considering the TESS point-spread-function with a full-width-half-maximum of approximately $40^{\prime \prime}$, and the typically irregularly shaped SPOC photometric apertures and circular Quick Look Pipeline \citep[QLP;][]{2020RNAAS...4..204H} photometric apertures, which usually extend to $\sim1^{\prime}$ from the target star, we extend our scrutiny to stars located up to $2.5^{\prime}$ away from the target star. In order to confidently clear a star of any nearby eclipsing binary (NEB) signal, we require that the light curve for that star exhibits a model residual RMS value that is at least three times smaller than the depth of the eclipse necessary to produce the TESS detection in that star. We also ensure that the predicted ephemeris uncertainty is covered by at least $\pm3\sigma$ relative to the most precise SPOC or QLP ephemeris available at the time of publication. Additionally, we manually inspect the light curves of all nearby stars to ensure the absence of any evident eclipse-like events. Through this meticulous process of elimination, we deduce that when all the requisite nearby stars are "cleared" of NEBs, the transit is indeed occurring on the target star, or on a star so close to the target star that it remained undetected by Gaia DR3 and is not listed in TIC version 8.

The VPC designation signifies that we have substantiated, through ground-based follow-up light curve photometry, that the TESS-detected event is unquestionably occurring on the intended target. This validation is achieved by employing follow-up photometric apertures of sufficiently reduced size, designed to exclude most or all of the flux originating from the nearest stars listed in Gaia DR3 and/or TIC version 8, which possess the requisite brightness to generate a signal similar to that observed by TESS.

The VPC+ classification is akin to VPC, with the added step of quantifying transit depths within the photometric apertures centered on the target star across various optical bands. We promote the disposition to VPC+ if no substantial transit depth discrepancy is observed among these bands, and such discrepancies do not exceed a significance level of more than 3 standard deviations ($>3\sigma$).

\subsubsection{LCOGT}
The 1.0\,m network nodes of the Las Cumbres Observatory Global Telescope \citep[LCOGT;][]{Brown:2013} are situated across various locations worldwide: Cerro Tololo Inter-American Observatory in Chile (CTIO), Siding Spring Observatory near Coonabarabran, Australia (SSO), South Africa Astronomical Observatory near Cape Town, South Africa (SAAO), Teide Observatory on the island of Tenerife (TEID), McDonald Observatory (MCD) near Fort Davis, Texas, United States, and Haleakala Observatory on Maui, Hawai'i (HAI). These telescopes feature $4096\times4096$ SINISTRO cameras, providing an image scale of $0.389^{\prime \prime}$ per pixel and offering a $26^{\prime}\times26^{\prime}$ field of view. Additionally, the LCOGT 2\,m Faulkes Telescope North at Haleakala Observatory is equipped with the MuSCAT3 multi-band imager \citep{Narita:2020}. Calibration of all LCOGT images was conducted using the standard LCOGT {\tt BANZAI} pipeline \citep{McCully:2018}, while differential photometric data were derived using {\tt AstroImageJ} \citep{Collins:2017}.

\subsubsection{M-Earth-South}
MEarth-South, established as detailed by \citet{Irwin:2007}, comprises eight 0.4\,m telescopes positioned at Cerro Tololo Inter-American Observatory, situated to the east of La Serena, Chile. These telescopes are outfitted with Apogee U230 detectors, providing a $29^{\prime}\times29^{\prime}$ field of view and an image scale of 0.84$^{\prime \prime}$ per pixel. Analysis of outcomes was conducted through custom pipelines expounded in \citet{Irwin:2007}.

\subsubsection{MuSCAT2}
The MuSCAT2 multi-color imager, detailed in \citet{Narita:2019}, is situated at the 1.52~m Telescopio Carlos Sanchez (TCS) within the Teide Observatory, Spain. MuSCAT2 conducts simultaneous observations in Sloan $g'$, Sloan $r'$, Sloan $i'$, and $z_S$. With an image scale of $0.44^{\prime \prime}$ per pixel, the imager provides a field of view measuring $7.4^{\prime}\times7.4^{\prime}$. Photometric analysis was executed utilizing standard aperture photometry calibration and reduction procedures via a dedicated MuSCAT2 photometry pipeline, elucidated in \citet{Parviainen:2020}.

\subsubsection{KeplerCam}
The 1.2\,m telescope at the Fred Lawrence Whipple Observatory, positioned on Mt. Hopkins in southern Arizona, incorporates the KeplerCam. Employing a $4096\times4096$ Fairchild CCD 486 detector, it yields an image scale of $0.672^{\prime \prime}$ per $2\times2$ binned pixel, thus delivering a field of view measuring $23^{\prime}.1\times23^{\prime}.1$. The image data underwent calibration, and photometric data were extracted through the utilization of {\tt AstroImageJ}.

\subsubsection{TRAPPIST}
The TRAnsiting Planets and PlanetesImals Small Telescope (TRAPPIST) North 0.6\,m telescope, situated at Oukaimeden Observatory in Morocco, and the TRAPPIST-South 0.6\,m telescope, located at La Silla Observatory near Coquimbo, Chile, are detailed in \citet{2011Msngr.145....2J, Gillon2011}. TRAPPIST-North, as expounded in \citet{Barkaoui2019_TN}, is outfitted with an Andor IKONL BEX2 DD camera, providing an image scale of 0.6$^{\prime \prime}$ per pixel, resulting in a $20^{\prime}\times20^{\prime}$ field of view. On the other hand, TRAPPIST-South utilizes an FLI camera, generating an image scale of 0.63$^{\prime \prime}.$ per pixel, resulting in a $22^{\prime}\times22^{\prime}$ field of view. Calibration of the image data and extraction of photometric data were performed using either {\tt AstroImageJ} or a dedicated pipeline leveraging the {\tt prose} framework delineated in \citet{Garcia:2022}.

\subsubsection{SPECULOOS-South}
The SPECULOOS Southern Observatory (SSO) \citep{2018Msngr.174....2J}, comprises four 1\,m telescopes, namely SSO-Io and SSO-Europa, located at the Paranal Observatory near Cerro Paranal, Chile. Equipped with detectors providing an image scale of 0.35$^{\prime \prime}$ per pixel, these telescopes offer a $12^{\prime}\times12^{\prime}$ field of view. Image data underwent calibration, and photometric data extraction was carried out using a specialized pipeline outlined in \citet{Sebastian:2020}.

\subsubsection{PEST}
The Perth Exoplanet Survey Telescope (PEST), situated near Perth, Australia, employed a 0.3 m telescope along with a $1530\times1020$ SBIG ST-8XME camera, offering an image scale of 1$^{\prime \prime}.$2 per pixel, resulting in a $31^{\prime}\times21^{\prime}$ field of view. Image calibration and the extraction of differential photometry were executed utilizing a customized pipeline based on {\tt C-Munipack}\footnote{\url{http://c-munipack.sourceforge.net}}.

\begin{table*}
    \centering
    \resizebox{\textwidth}{!}
{
    \begin{tabular}{c c c c l c}
\hline
TOI & Observatory & UTC Date & Filter & \multicolumn{1}{c}{Results} & Disp.\footnotemark[1] \\
\hline
\hline
TOI-238.01  & LCO-SAAO  & 2019-06-15 & $i'$ & cleared the field of NEBs &  \\
            & LCO-TEID  & 2022-11-28 & $z_s$\footnotemark[2] & cleared the field of NEBs  & CPC\\
            \hline
TOI-771.01  & M-Earth-South & 2020-02-02 & RG715\footnotemark[3] & $\sim3$ ppt transit in 5$^{\prime \prime}$ target aperture &   \\
            & TRAPPIST-S & 2022-01-29  & I+z   & $\sim3$ ppt transit in 4.5$^{\prime \prime}$ target aperture &  \\
            & SSO-Io     & 2022-02-05  & $g'$   & $\sim3$ ppt transit in 3.3$^{\prime \prime}$ target aperture &  \\
            & TRAPPIST-S & 2022-02-12  & $z'$   & $\sim3$ ppt transit in 5$^{\prime \prime}$ target aperture &  \\
            & SSO-Europa & 2022-04-02  & $g'$   & $\sim3$ ppt transit in 3.6$^{\prime \prime}$ target aperture & VPC+ \\
            \hline
TOI-871.01  & PEST       & 2020-01-27  & Rc     & no detection on-target, cleared field of NEBs &   \\
            & LCO-SSO    & 2021-12-17  & $i',z'$& tentative $\sim0.8$ ppt target in 5.1$^{\prime \prime}$ target aperture & VPC \\
            & LCO-CTIO   & 2023-10-09  & $z_s$  & $\sim0.9$ ppt transit using an uncontaminated 6.2$^{\prime \prime}$ target aperture. &\\
            \hline
TOI-1467.01 & MuSCAT3   & 2021-07-28  & $i',z_s$& turned out to be out-of-transit &   \\
             & LCO-TEID  & 2021-09-14  & $z_s$  & $\sim1.4$ ppt transit in 5.9$^{\prime \prime}$ target aperture&  \\
             & MuSCAT3   & 2022-08-22  & $g'r'i'z_s$& $\sim1.4$ ppt transit in 4.5$^{\prime \prime}$ target aperture & VPC+ \\
             \hline
TOI-1739.01 & LCO-McD   & 2020-05-06  & $z_s$ & $\sim0.6$ ppt transit in 7.8$^{\prime \prime}$ target aperture & \\
             & LCO-McD   & 2022-05-15  & $r'$  & $\sim1$ ppt transit in 8.7$^{\prime \prime}$ target aperture &  \\
             & LCO-TEID  & 2022-05-31  & $r'$  & $\sim1$ ppt transit in 12.9$^{\prime \prime}$ target aperture&  VPC \\
             \hline
TOI-2068.01 & LCO-McD   & 2020-12-20  & $i'$  & $\sim1.5$ ppt transit in 8.2$^{\prime \prime}$ target aperture, cleared field of NEBs& \\
             & KeplerCam & 2021-04-08  & $i'$  & $\sim1$ ppt transit in 6.7$^{\prime \prime}$ target aperture &  \\
             & LCO-TEID  & 2022-04-16  & $i'$  & $\sim1.3$ ppt transit in 5.1$^{\prime \prime}$ target aperture &  \\
             & MuSCAT2   & 2023-07-02  & $i'z_s$& $\sim$ cleared 2 of the brightest nearby stars for NEBs& VPC  \\
             \hline
TOI-4559.01 & LCO-SSO   & 2022-04-18  & $i'$  & $\sim1.3$ ppt transit in 3.5$^{\prime \prime}$ target aperture, cleared field of NEBs& \\
             & LCO-CTIO  & 2022-06-25  & $i'$  & $\sim2$ ppt transit in 5.9$^{\prime \prime}$ target aperture &  \\
             & LCO-CTIO   & 2022-07-02 & $i'$   & $\sim1.2$ ppt transit 5$^{\prime \prime}$ target apertures&  VPC  \\
             \hline
TOI-5799.01 & TRAPPIST-N & 2022-09-28 & $z'$   & cleared field of NEBs &   \\
             & LCO-TEID   & 2023-06-14 & $i'$ &$\sim2.6$ ppt transit in 3.9$^{\prime \prime}$ target aperture &  \\
             & MuSCAT2    & 2023-07-08 & $g'r'i'$& $\sim3$ ppt transit in 10.8$^{\prime \prime}$ target aperture& VPC+ \\
\hline
    \end{tabular}
    }
    \footnotemark[1]{The overall follow-up disposition. CPC $=$ cleared of NEBs, VPC $=$ on-target relative to Gaia DR3 stars, VPC+ $=$ achromatic on-target relative to Gaia DR3 stars. See the text for full disposition definitions.}
    \footnotemark[2]{Pan-STARRS $z$-short band ($\lambda_{\rm c} = 8700$\,\AA, ${\rm Width} =1040$\,\AA)}
    \footnotemark[3]{7150\AA long-pass filter}
    \caption{Ground-based light curve observations.}
    \label{tab:transitfollowup_VaTESTIII}
\end{table*}

\subsection{High Resolution Imaging}
Employing high-resolution imaging techniques like adaptive optics and speckle imaging significantly minimizes the likelihood of blended background objects. TFOP Sub Group 3 (SG3), collected the data, with specifics detailed in Table \ref{tab:HRI_data}, visually represented in Figure \ref{fig:CC_Plots}, and thoroughly expounded upon in the subsequent sections.

\subsubsection{Gemini-N/'Alopeke, and Gemini-S/Zorro}
\begin{figure*}
    \centering
    \includegraphics[scale = 0.43]{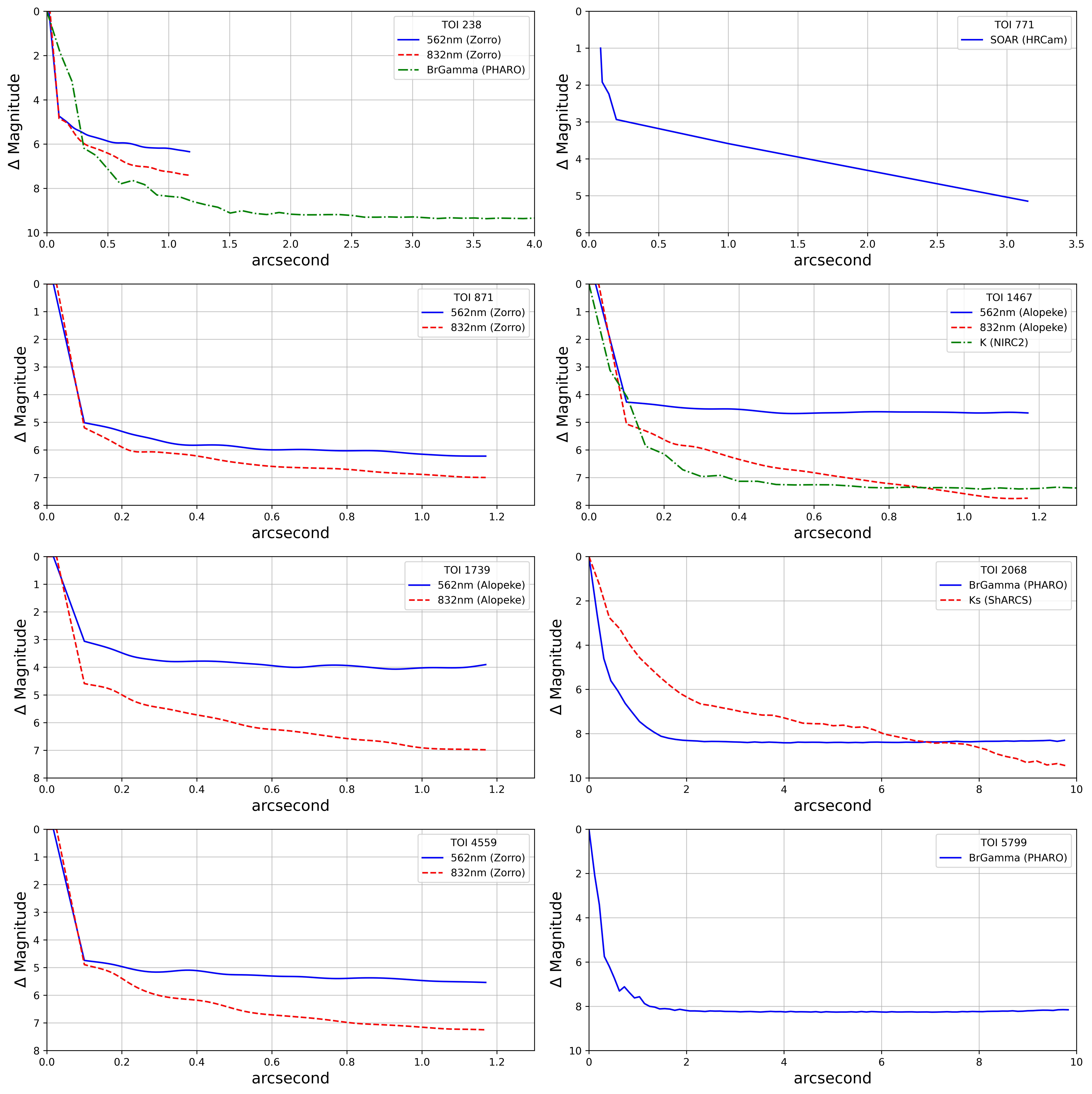}
    \caption{The contrast curves derived from the high-resolution follow-up observations enable us to eliminate the possibility of companions at specific separations beyond a certain magnitude difference ($\Delta$ Magnitude).}
    \label{fig:CC_Plots}
\end{figure*}

{
'Alopeke and Zorro, positioned at the calibration ports of Gemini North and South, conducted speckle interferometric measurements, detailed in \citep{HorchAlopke2009AJ....137.5057H,Scottalopke2021FrASS...8..138S}. The complete image datasets for each star at 562~nm ($\Delta \lambda$ = 54~nm) and 832~nm ($\Delta \lambda$ = 40~nm) were merged and analyzed in Fourier space to derive their power spectrum and auto-correlation functions, as outlined in \citep{2011AJ....142...19H}. The culmination of the data reduction process generated 5$\sigma$ contrast curves in each filter, which set constraints on any potential companions located in very close proximity to the different candidates. Figure \ref{fig:CC_Plots} exhibits the contrast curves acquired for each star. Examination of the Fourier analysis unveiled the absence of nearby secondary sources.}

\subsubsection{Keck/NIRC2}{
High-resolution imaging observations was utilized using NIRC2 \citep{Sakai2019ApJ...873...65S} positioned on Keck-II's left Nasmyth Platform \citep{Wizinowich2000PASP..112..315W}, behind the Adaptive Optics (AO) bench. We adhered to the observation plan and analysis method outlined in \citet{Schlieder2021FrASS...8...63S} for conducting high-resolution imaging of TESS systems with the NIRC2 instrument. In summary, our observations involved 0.181-second integrations following a standard dither sequence consisting of 3$\prime \prime$ steps, repeated thrice, with each subsequent dither offset by 0.5$\prime \prime$. At each position, we performed 1 co-add, resulting in a total of 9 frames. We utilized the narrow-angle mode of the NIRC2 camera, characterized by a plate scale of 9.942 milliarcseconds per pixel and a 10$\prime \prime$ field of view. Employing simulated sources at discrete separations, incrementally varied azimuthally at $45^\circ$ intervals and set at integer multiples of the central source's full width at half maximum (FWHM), we gauged sensitivity to nearby stars \citep{Schlieder2021FrASS...8...63S}. Contrast sensitivity was determined by raising the flux of each simulated source until aperture photometry detected a signal at $5\sigma$. Averaging all the limits at that separation yielded the final contrast sensitivity relative to separation. TOI-1467 observations utilized the Ks filter $(\lambda_0$ = 2.146$\mu$m; $\Delta\lambda$ = 0.311$\mu$m). Insights from Keck AO observations revealed no additional stellar companions.}\label{sec:Keck/NIRC2}

\subsubsection{Palomar/PHARO}{
PHARO, detailed in \citep{Hayward2001PASP..113..105H}, is a near-infrared camera tailored for operation alongside the Palomar Observatory's 200-inch Hale telescope and the Palomar Adaptive Optics system. Its detector comprises a $1024\times1024$ Rockwell HAWAII HgCdTe pixel array, sensitive within the wavelength range of 1-1.25 $\mu$m. This setup achieves diffraction-limited angular resolutions of 0.063 and 0.111 for J and K band imaging, respectively. Featuring a large-format detector, it encompasses a field of view spanning 25 to 40 degrees. PHARO was employed for AO imaging of TOI-238, TOI-2068, and TOI-5799 in Br$\gamma$ $(\lambda_0$ = 2.166$\mu$m, $\Delta\lambda$= 0.02$\mu$m). Estimated contrasts at various separations are presented in Table \ref{tab:HRI_data}. No secondary sources were identified in the reconstructed images.}

\subsubsection{Shane/ShARCS}{
The ShARCS camera, stationed at the Lick Observatory's Shane 3-meter telescope \citep{Kupke2012SPIE.8447E..3GK,McGurk2014SPIE.9148E..3AM}, utilized the Shane adaptive optics system in natural guide star mode to search for nearby, unresolved stellar companions. Observation sequences were obtained employing the $K_S$ filter $(\lambda_0 = 2.150 \mu$m, $\Delta\lambda= 0.320 \mu$m). Data reduction was conducted using the publicly available \texttt{SImMER} pipeline, outlined in \citep{Savel2022PASP..134l4501S}. In the case of TOI-2068, our observations, at $1^{\prime \prime}$, achieved a contrast of 4.455 (Ks). Within the scope of our detection limits, no neighboring stellar companions were identified.}

\subsubsection{SOAR/HRCam}{
We conducted speckle imaging observations for TOI-771 utilizing the high-resolution camera (HRCam), capable of observing a $9.9^{\prime\prime} \times 7.5^{\prime\prime}$ field of the sky through a $658\times 496$-pixel array. Each pixel captures light from a 15 milli-arcsecond region \citep{Tokovinin2010AJ....139..743T}. This instrument, designed for swift imaging, is intended for use with the SOAR telescope and employs a CCD detector featuring built-in electro-multiplication. Further details and the subsequent analysis are extensively discussed in literature cited as \citep{Zieglar2020AJ....159...19Z,Zieglar2021AJ....162..192Z}. For a deeper understanding, we recommend referring to those articles. Insights from SOAR AO observations revealed no additional stellar companions.}

\subsubsection{SAI/SPeckle Polarimeter}
We observed TOI-2068 on 2020 November 29 UT with the SPeckle Polarimeter (SPP) \citep{Safonov2017AstL...43..344S} on the 2.5~m telescope at the Caucasian Observatory of Sternberg Astronomical Institute (SAI) of Lomonosov Moscow State University. Electron Multiplying CCD Andor iXon 897 was employed as a detector. The atmospheric dispersion compensator was active. Observations were conducted in the $I_c$ band. The power spectrum was estimated from 4000 frames with 30 ms exposure. The detector has a pixel scale of $20.6$ mas pixel$^{-1}$, and the angular resolution was 83 mas. Field of view is $5^{\prime\prime}\times10^{\prime\prime}$. We did not detect any stellar companions; limits of detection are given in Table~\ref{tab:HRI_data}.

\begin{table*}
\centering
{
\begin{tabular}{ccccccccccc}
\hline
TOI & 
Observation Date (UT) &
Telescope & 
Instrument &
Filter &
Image Type & 
\multicolumn{5}{c}{Contrast $\Delta$mag} \\
\cmidrule{7-11}
 & 
 &
 & 
 &
 & 
 &
0.1$^{\prime\prime}$ &
0.5$^{\prime\prime}$ &
1.0$^{\prime\prime}$ &
1.5$^{\prime\prime}$ & 
2.0$^{\prime\prime}$\\ 
\hline
\hline
238 & 2019-09-12 & Gemini-S (8m) & Zorro & 562 nm & Speckle & 4.73 & 5.86 & 6.19 & --- & --- \\
    & 2019-09-12 & Gemini-S (8m) & Zorro & 832 nm & Speckle & 4.83 & 6.41 & 7.25 & --- & --- \\
    & 2019-07-14 & Palomar (5m) & PHARO & BrGamma & AO & 1.87 & 7.15 & 8.36 & 9.11 & 9.17 \\
\hline
771 & 2019-07-14 & SOAR (4.1 m) & HRCam & I & Speckle & 1.92 & --- & 3.58 & --- & --- \\
\hline
871 & 2020-12-28 & Gemini-S (8m) & Zorro & 562 nm & Speckle & 5.02 & 5.87 & 6.15 & --- & --- \\
    & 2020-12-28 & Gemini-S (8m) & Zorro & 832 nm & Speckle & 5.19 & 6.44 & 6.88 & --- & --- \\
\hline
1467 & 2021-10-18 & Gemini-N (8m) & 'Alopeke & 562 nm & Speckle & 4.27 & 6.35 & 4.65 & --- & --- \\
     & 2021-10-18 & Gemini-N (8m) & 'Alopeke & 832 nm & Speckle & 5.06 & 6.65 & 7.58 & --- & --- \\
     & 2020-09-09 & Keck2 (10m) & NIRC2 & K & AO & 4.09 & 7.25 & 7.37 & 7.36 & 7.36 \\
\hline
1739 & 2020-06-08 & Gemini-N (8m) & 'Alopeke & 562 nm & Speckle & 3.06 & 3.83 & 4.02 & --- & --- \\
     & 2020-06-08 & Gemini-N (8m) & 'Alopeke & 832 nm & Speckle & 4.58 & 6.01 & 6.91 & --- & --- \\
\hline
2068 & 2021-06-21 & Palomar (5m) & PHARO & BrGamma & AO & 1.56 & 5.77 & 7.45 & 8.13  & 8.31 \\
     & 2020-12-01 & Shane (3m) & ShARCS & Ks & AO & 0.59 & 2.95 & 4.45 & 5.53 & 6.34 \\
     & 2020-11-29 & SAI (2.5m) & SPP    & I  & Speckle & 2.11 & 5.23 & 5.83 & --- & --- \\
\hline
4559 & 2022-03-17 & Gemini-S (8m) & Zorro & 562 nm & Speckle & 4.74 & 5.26 & 5.47 & --- & --- \\
     & 2022-03-17 & Gemini-S (8m) & Zorro & 832 nm & Speckle & 4.89 & 6.49 & 7.15 & --- & --- \\
\hline
5799 & 2023-07-01 & Palomar (5m) & PHARO & BrGamma & AO & 1.77 & 6.62 & 7.58 & 8.11 & 8.18 \\
\hline
\end{tabular}
}
\caption{Details of High-resolution Imaging data. \label{tab:HRI_data}}
\end{table*}

\section{Statistical Validation using \texttt{TRICERATOPS}}\label{sec:triceratops}
\texttt{TRICERATOPS} \citep{2020ascl.soft02004G,2021AJ....161...24G} is one of the widely used statistical tools to validate exoplanets. It makes use of the Bayesian framework starting by searching for background stars within a specific radius (2.5$^{\prime \prime}$) of the target to determine the contamination of the flux due to these stars. Next, based on the contamination in the flux, \texttt{TRICERATOPS} calculates the probability of that signal being generated by a transiting planet, an eclipsing binary, or a nearby eclipsing binary. This is done using Marginal Likelihood and is combined with the prior to estimate Nearby False Positive Probability (NFPP) and False Positive Probability (FPP) which is given by

\begin{eqnarray}
    NFPP &=& \sum (\mathcal{P}_{NTP} + \mathcal{P}_{NEB} + \mathcal{P}_{NEBX2P}) \\
    FPP &=& 1 - (\mathcal{P}_{TP} + \mathcal{P}_{PTP} + \mathcal{P}_{DTP})
\end{eqnarray}

Here $\mathcal{P}_j$ is the probability of each scenario that can be found in Table 1 of \citet{2021AJ....161...24G}, i.e., TP = No unresolved companion; transiting planet with Period around target star, PTP = Unresolved bound companion; transiting planet with Period around a primary star, DTP = Unresolved background star; transiting planet with Period around target star, NTP = No unresolved companion; transiting planet with Period around the nearby star, NEB = No unresolved companion; eclipsing binary with Period around the nearby star and NEBX2P = No unresolved companion; eclipsing binary with 2 $\times$ Period around a nearby star. Please refer to \citet{2021AJ....161...24G} for more detailed information.

\texttt{TRICERATOPS} assumes that the user has visually inspected the light curves for obvious signatures of astrophysical false positives, such as secondary eclipses and odd-even transit depth differences, and has ruled them out. For this, we refer to the TESS SPOC DV reports \citep{2016SPIE.9913E..3EJ}, which conducts an automated search for these features. For each of the TOIs analyzed here, the corresponding DV report finds no strong evidence of a secondary eclipse or depth variations between odd-numbers and even-numbers transits. \texttt{TRICERATOPS} also assumes that the planet candidate is not a false alarm originating from stellar activity or instrumental sources. We visually inspected each of the light curves and found no evidence of stellar activity with periods matching the TOIs we analyze. In addition, the transit times and orbital periods of the TOIs do not match the cadence of TESS momentum dumps, which are common sources of instrumental false alarms \citep[e.g.,][]{2023RNAAS...7....7K}. Lastly, we note that the transits are persistent in the TESS data (i.e., there are no mysteriously missing signals) and have morphologies consistent with an astrophysical origin. We therefore conclude that these TOIs are unlikely to be false alarms.

As discussed in Section \ref{sec:Ground_Based_Phtotmetry}, for our selected candidates, we already cleared the nearby stars that could be contaminating the transit signal, and confirmed the source of the signal on-target. So in our \texttt{TRICERATOPS} FPP calculations we did not use any nearby stars and thus the NFPP will be zero for all the candidates. Excluding the Nearby False Positive scenarios, \texttt{TRICERATOPS} tests for the presence of the following false positives: (1) the target is actually a double- or triple-star system where one of the companions eclipses the primary component, (2) the target is actually a hierarchical star system with a pair of eclipsing binary stars orbiting far from the primary component, (3) the target is a double-star system where the secondary component hosts a transiting planet, (4) there is a pair of chance-aligned foreground or background eclipsing binary stars, and (5) there is a chance-aligned foreground or background star that hosts a transiting planet. As per the threshold provided for \texttt{TRICERATOPS}, we considered the candidates as planets if FPP is 0.01 or smaller and NFPP less than 0.001 \citep{2021AJ....161...24G}. Results of \texttt{TRICERATOPS} FPP calculations are detailed in Table \ref{tab:FPP}, confirming that all the selected candidates are validated planets. Notably, our analysis revealed no instances of false positives. \texttt{TRICERATOPS} simulations are uploaded on \texttt{GitHub} repository\footnote{\url{https://github.com/supremeKAI40/Vatest-3-SuperEarth-statistical-validation.git}}.

\begin{table}
    \centering
    \begin{tabular}{c c c}
    \hline
    TOI ID & Contrast Curve\footnotemark[1] & False Positive Probability \\
    \hline
    \hline
    & & \\
    TOI 238 & PHARO (BrGamma) & (2.48 $\pm$ 0.17) $\times$ 10$^{-3}$ \\
            & Zorro (562nm) & (2.50 $\pm$ 0.17) $\times$ 10$^{-3}$ \\
            & Zorro (832nm) & (2.08 $\pm$ 0.11) $\times$ 10$^{-3}$\\
    \hline
    & & \\
    TOI 771 & HRCam (I) & (2.28 $\pm$ 5.30) $\times$ 10$^{-5}$ \\
    \hline
    & & \\
    TOI 871 & Zorro (562nm) & (1.59 $\pm$ 0.41) $\times$ 10$^{-3}$ \\
            & Zorro (562nm) & (1.17 $\pm$ 0.27) $\times$ 10$^{-3}$ \\
    \hline
    & & \\
    TOI 1467 & 'Alopeke (562nm) & (1.21 $\pm$ 0.46) $\times$ 10$^{-5}$ \\
             & 'Alopeke (832nm) & (5.03 $\pm$ 2.14) $\times$ 10$^{-6}$ \\
             &  NIRC2 (K) & (4.04 $\pm$ 2.12) $\times$ 10$^{-6}$ \\
    \hline
    & & \\
    TOI 1739 & 'Alopeke (562nm) & (5.22 $\pm$ 0.99) $\times$ 10$^{-3}$ \\
             & 'Alopeke (832nm) & (1.87 $\pm$ 0.23) $\times$ 10$^{-3}$ \\
    \hline
    & & \\
    TOI 2068 & PHARO (BrGamma) & (1.63 $\pm$ 0.15) $\times$ 10$^{-3}$ \\
             & ShARCS (Ks) & (8.57 $\pm$ 0.48) $\times$ 10$^{-3}$ \\
    \hline
    & & \\
    TOI 4559 & Zorro (562nm) & (9.61 $\pm$ 1.81) $\times$ 10$^{-3}$ \\
             & Zorro (832nm) & (3.56 $\pm$ 0.56) $\times$ 10$^{-3}$ \\
    \hline
    & & \\
    TOI 5799 & PHARO (BrGamma)  & (3.55 $\pm$ 1.08) $\times$ 10$^{-5}$ \\
    \hline
    \end{tabular} \\
    \footnotemark[1]{Details of contrast curve (high resolution imaging) used.}
    \caption{FPP calculated using \texttt{TRICERATOPS}.}
    \label{tab:FPP}
\end{table}

\section{TESS Data Reduction \& Modeling Techniques}\label{sec:Characteristics of the Validated Systems}
In this paper, we validate a total of 8 potential super-Earths. We make use of \texttt{Lightkurve} \citep{2018ascl.soft12013L} to download the data and \texttt{Juliet} \citep{2018ascl.soft12016E, 2019MNRAS.490.2262E} to model them and derive the planetary and orbital parameters. 

We download transit photometry data from the Mikulski Archive for Space Telescopes (MAST)\footnote{\url{https://archive.stsci.edu/}} using \texttt{Lightkurve} \citep{2018ascl.soft12013L} Python package. These light curves had stellar variability, which was removed by de-trending them with the \texttt{flatten} function of \texttt{Lightkurve}. We masked the in-transit portion of a signal during de-trending to ensure that the in-transit portion of the signal was not lost. To determine the characteristics of the transit event, including the transit duration, the epoch time, and the transit depth, we employed the Transit Least Squares (\texttt{TLS}) model, as outlined by \citep{2019A&A...623A..39H}. This analysis was conducted on the light curve data following the application of cleaning and detrending procedures. It's worth noting that the values provided by the Space Photometry Operation Center (SPOC) on the ExoFOP-TESS{https://exofop.ipac.caltech.edu/tess/} website are consistent with the results calculated using the \texttt{TLS} model.

\begin{figure*}
    \centering
    \includegraphics[scale = 0.31]{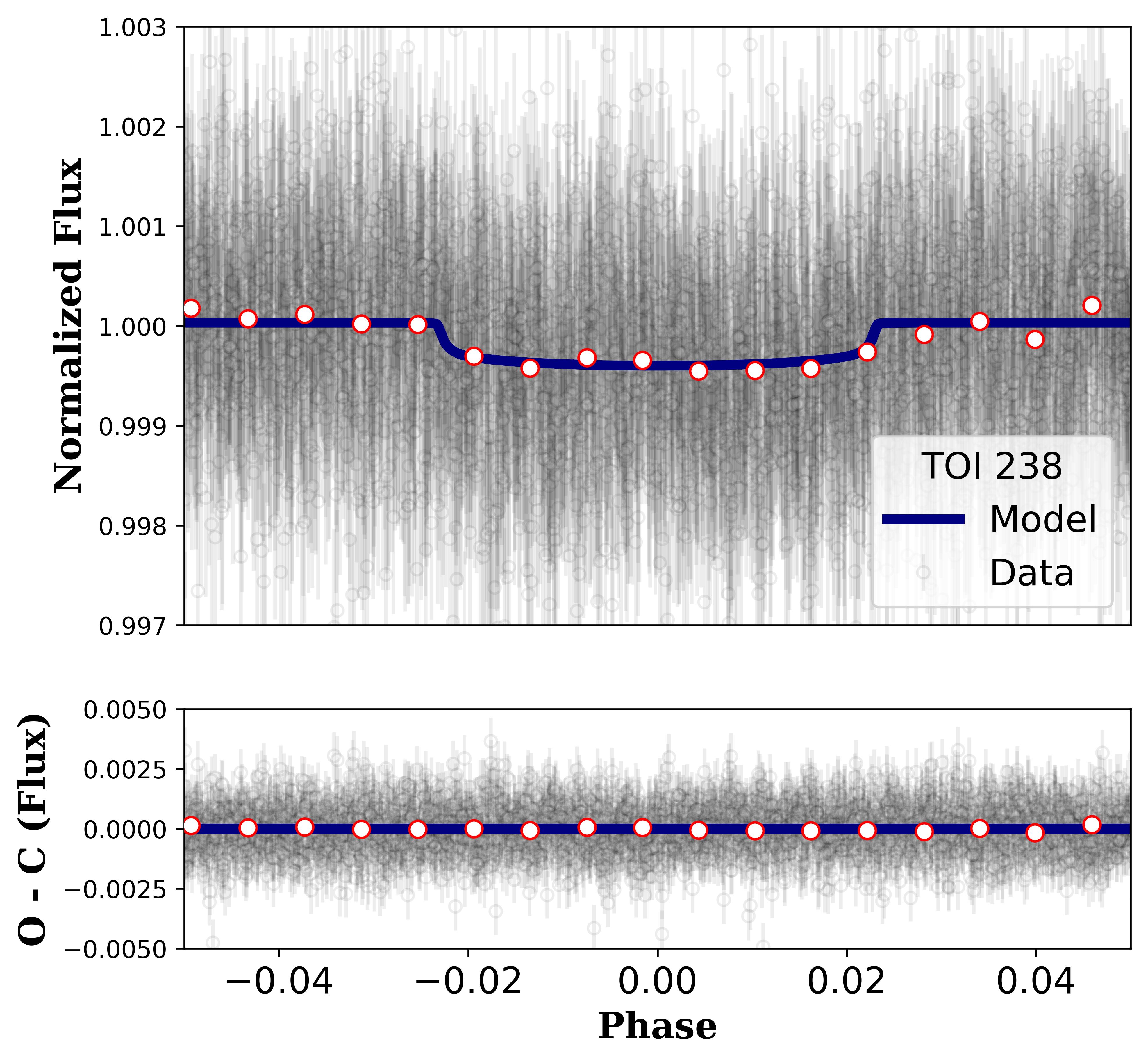}
    \includegraphics[scale = 0.31]{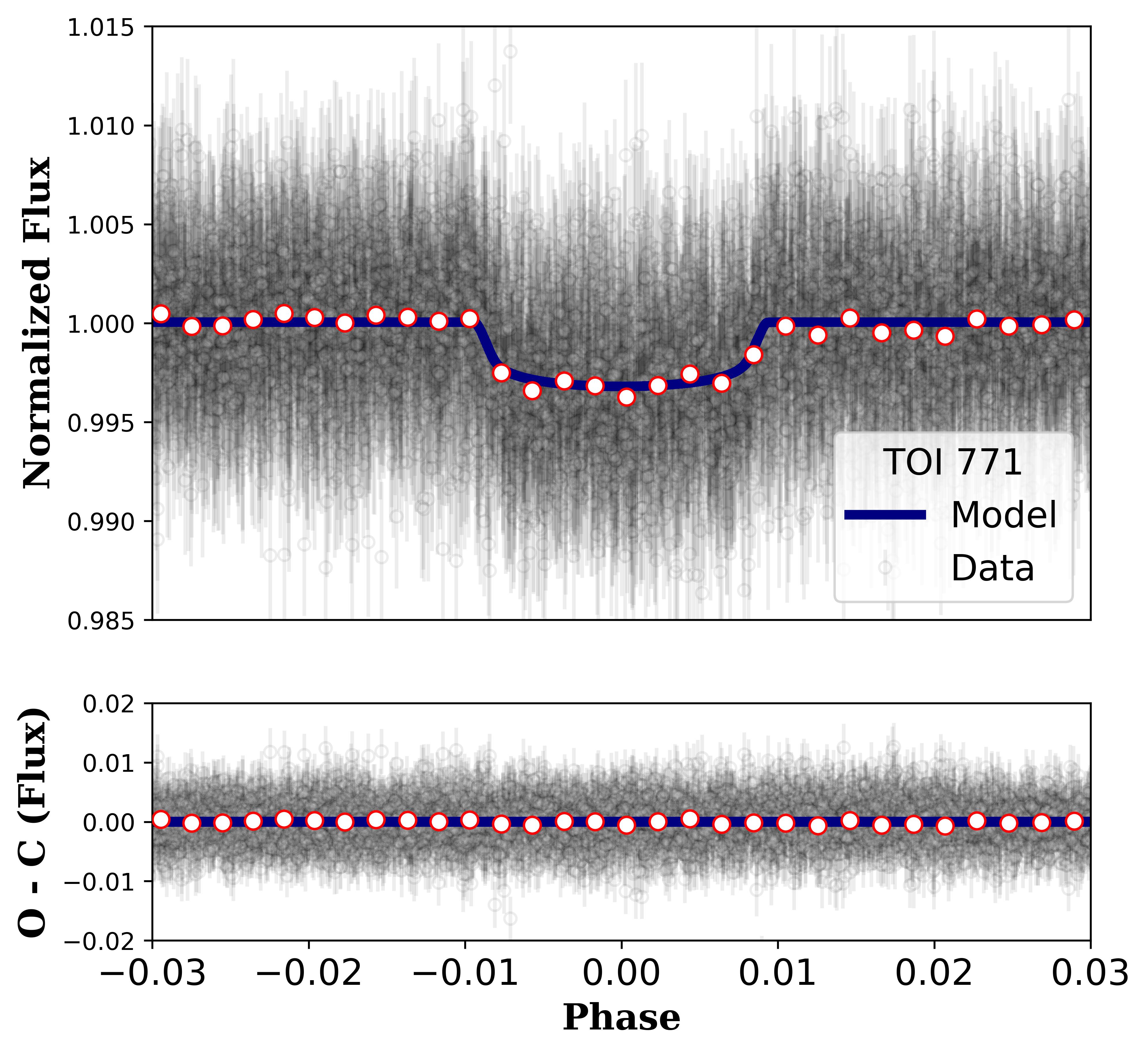}
    \includegraphics[scale = 0.31]{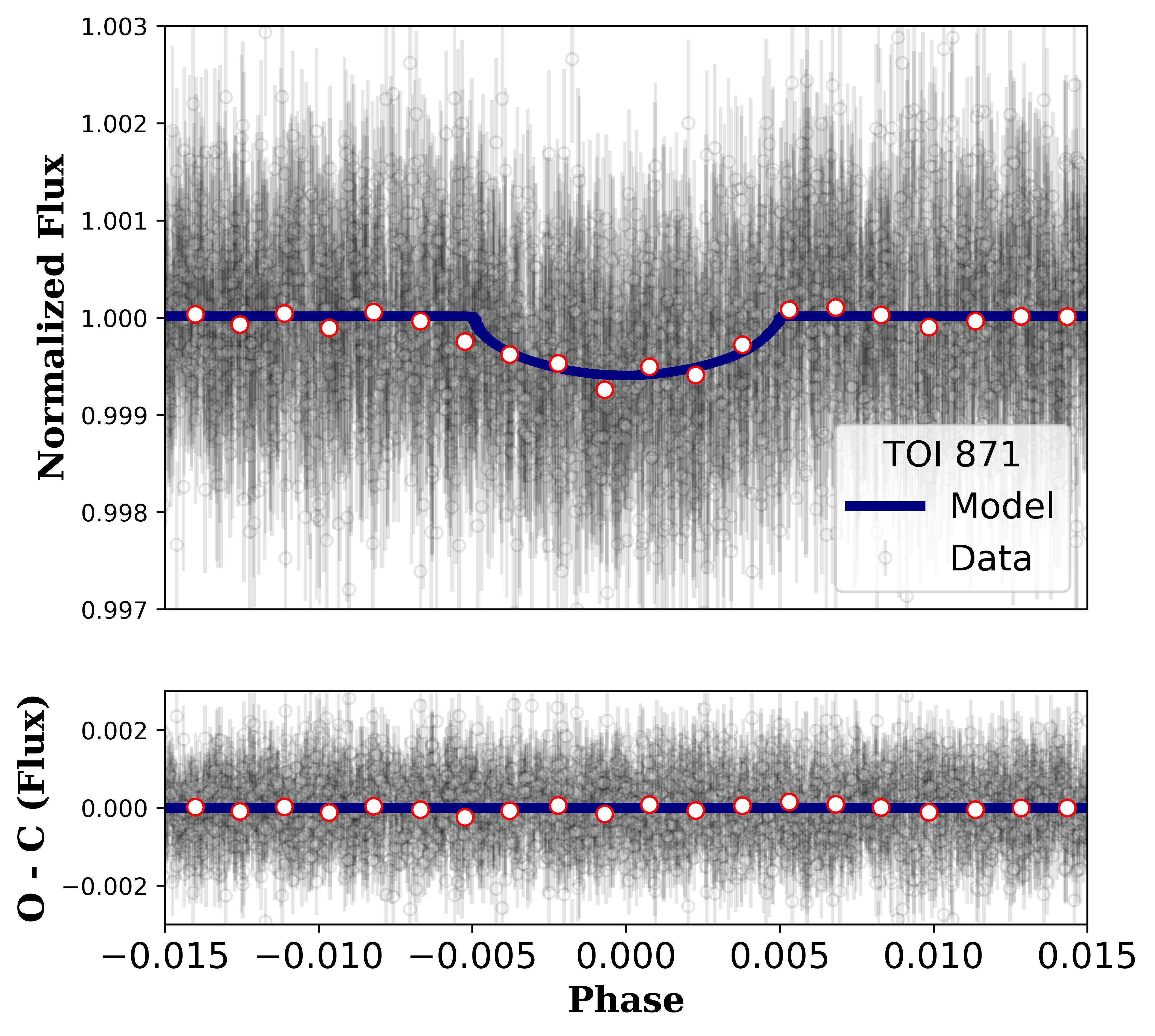}
    \includegraphics[scale = 0.31]{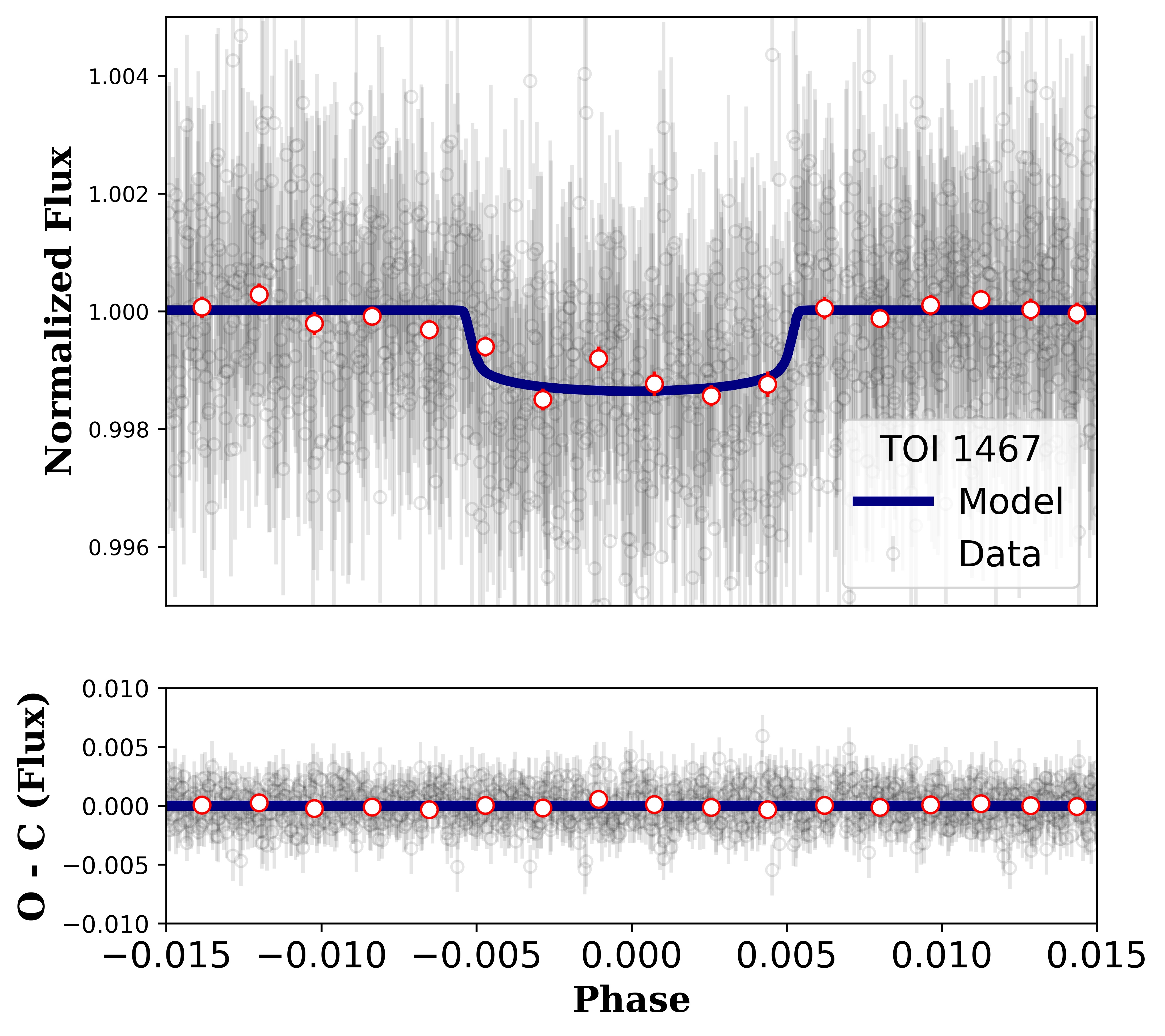}
    \includegraphics[scale = 0.31]{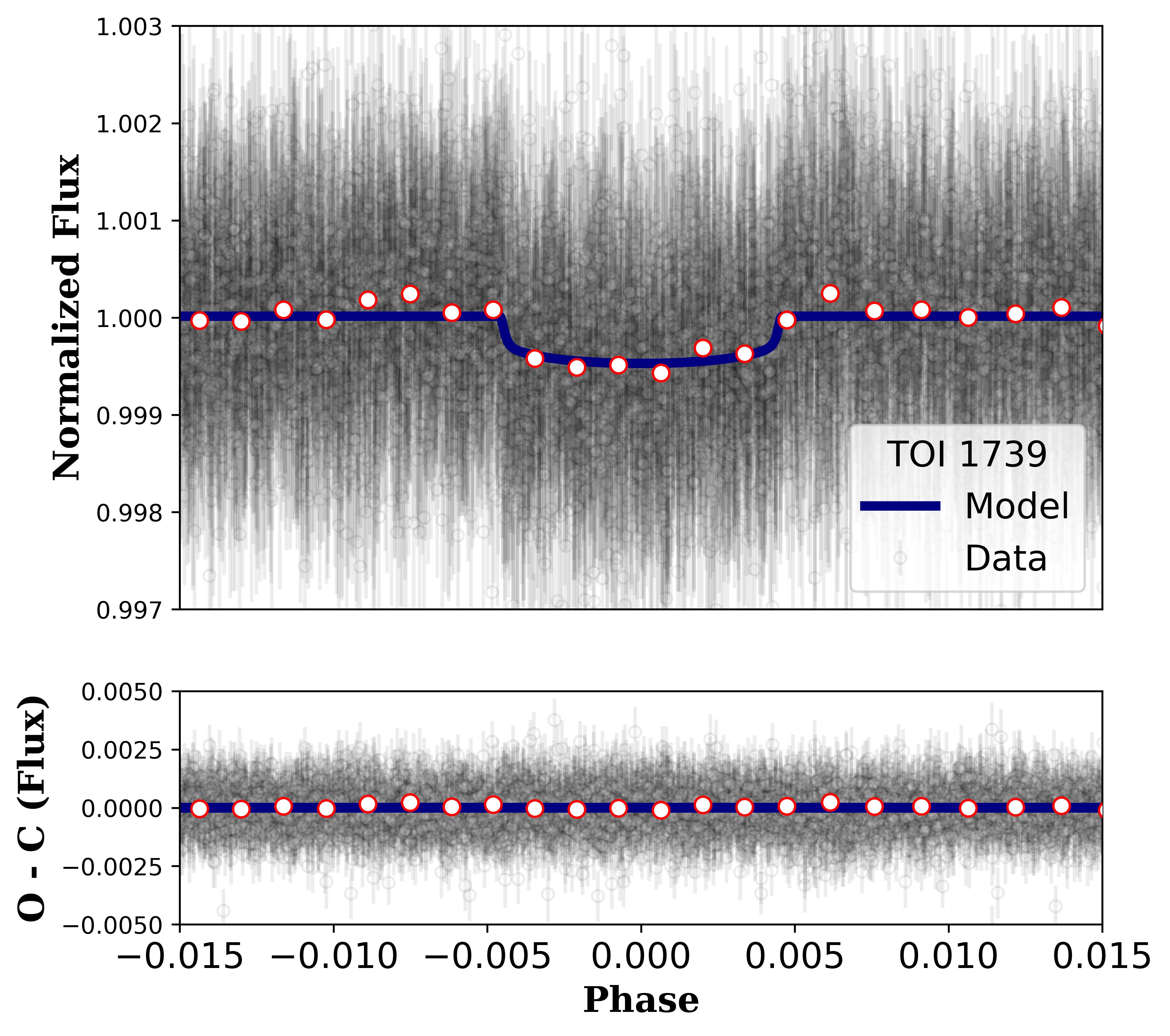}
    \includegraphics[scale = 0.31]{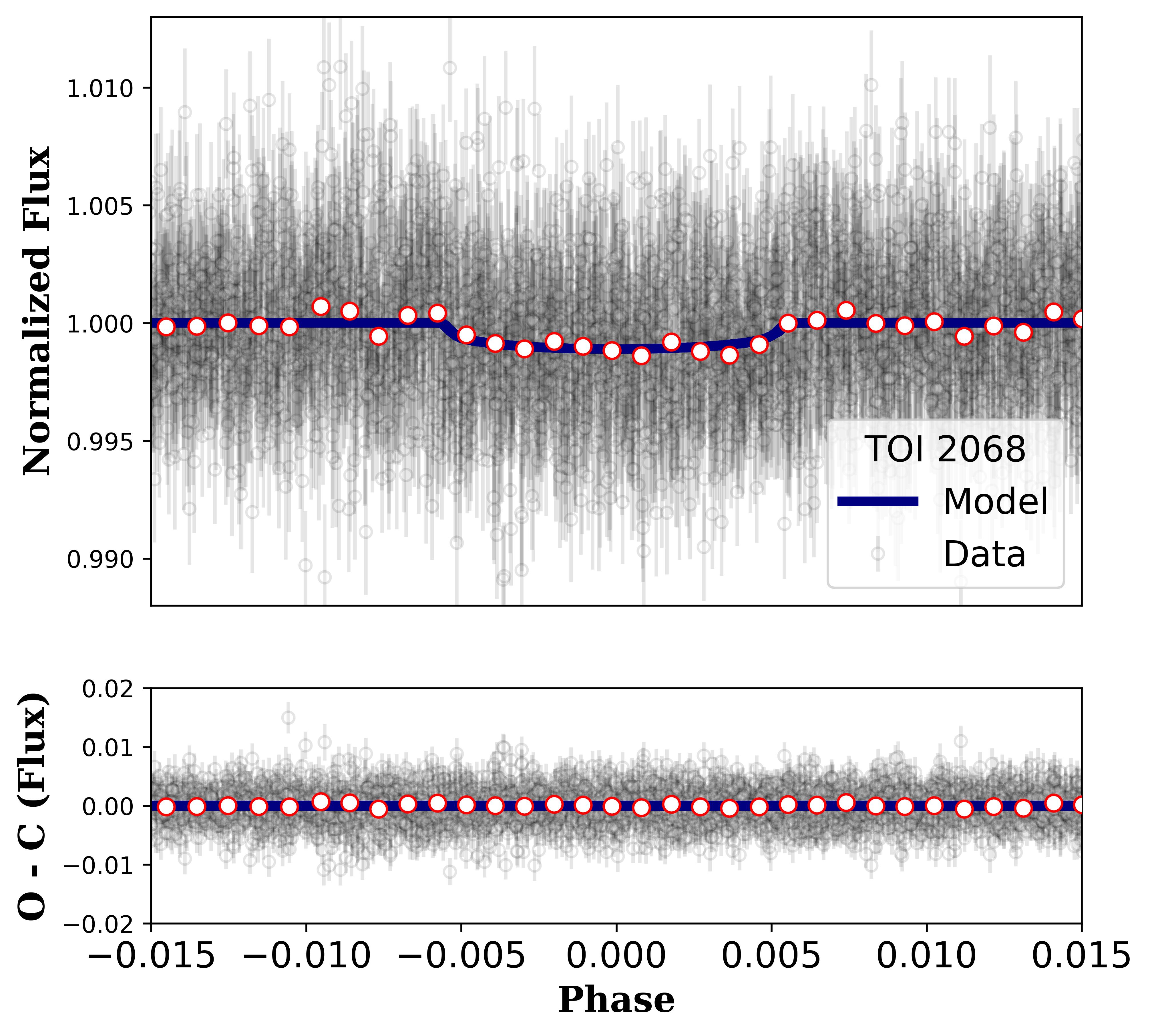}
    \includegraphics[scale = 0.31]{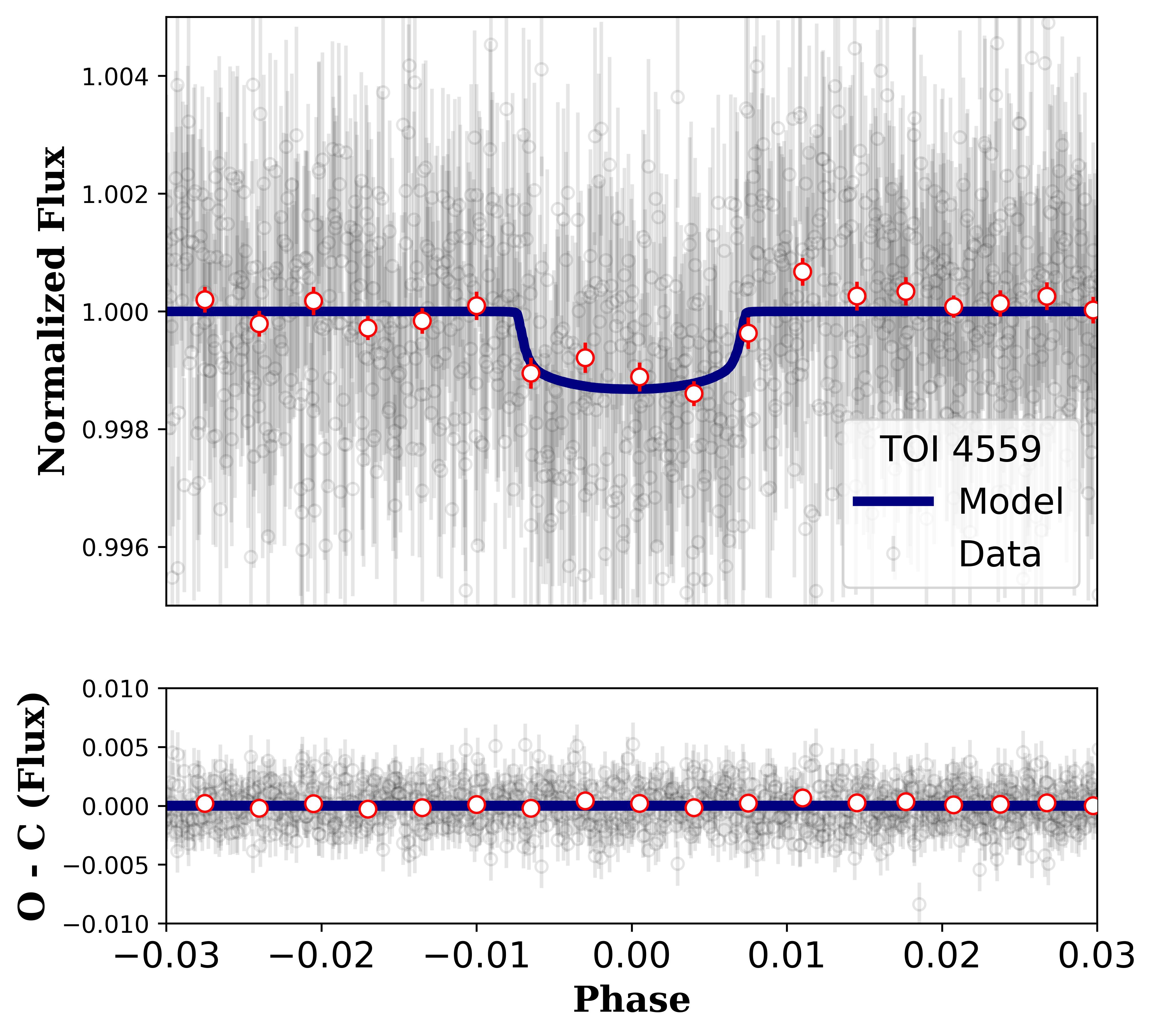}
    \includegraphics[scale = 0.31]{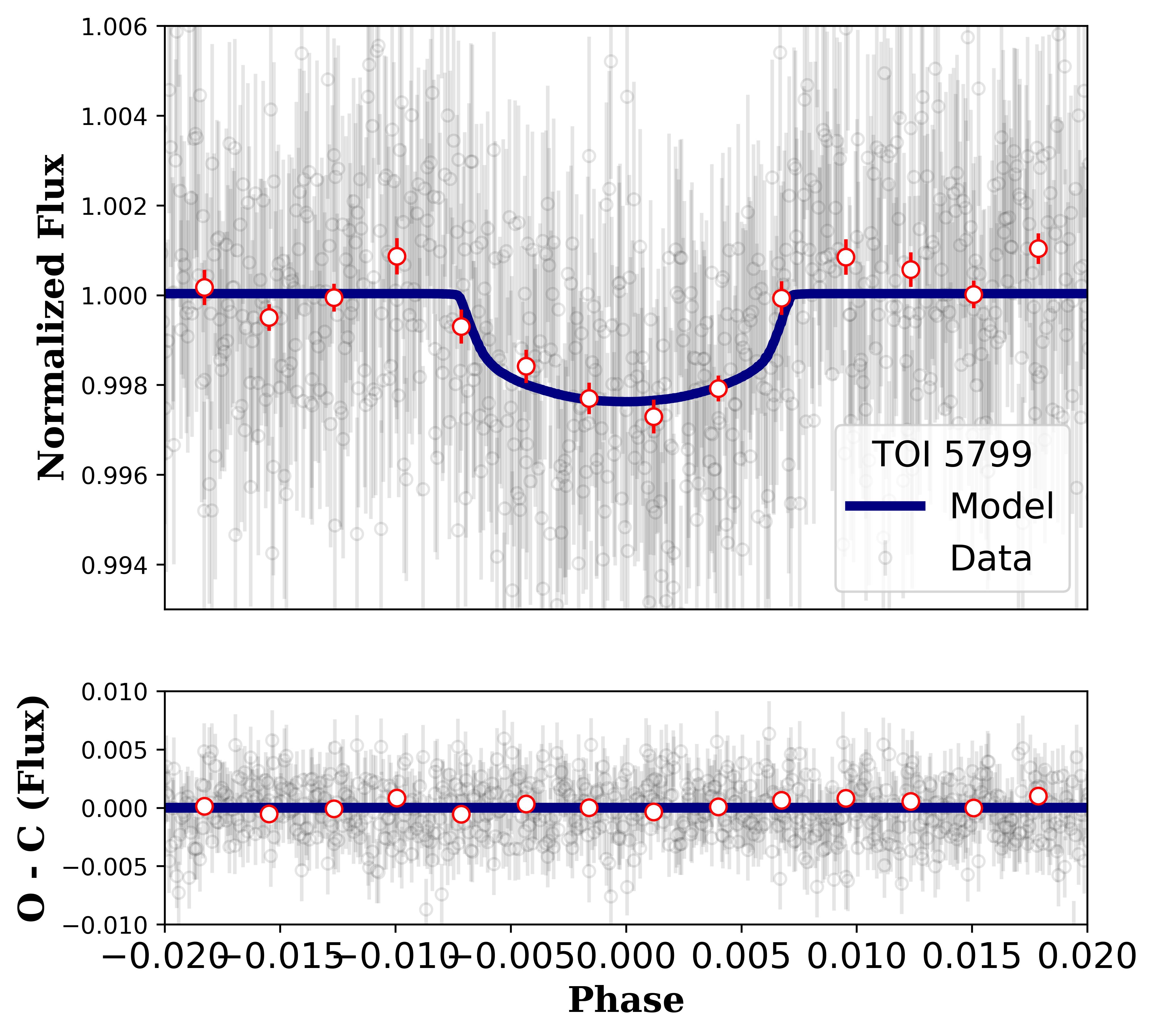}
    \caption{The figure displays phase-folded light curves for recently validated planetary systems. The dark blue line represents the optimal fitting model, while the red dots represent binned observations. The gray error bars in the background are data obtained by TESS.}
    \label{fig:transit_juliet}
\end{figure*}

The data was modeled using the \texttt{Juliet} transit modeling tool, as outlined in \citep{2018ascl.soft12016E, 2019MNRAS.490.2262E}. \texttt{Juliet} serves as a versatile tool for modeling exoplanetary systems, enabling the rapid and straightforward computation of parameters based on transit photometry, radial velocity, or both through Bayesian inference. It utilizes Nested Sampling for effective measurement and model comparisons. \texttt{Juliet} accommodates various datasets of transit photometry and radial velocity concurrently, calculating systematic trends with linear models or Gaussian Processes (GP). The priors employed during the modeling process are detailed in Table \ref{tab:priors}. For modeling TESS photometry data, we employed the \texttt{dynesty} sampler within the \texttt{Juliet} tool, and the resulting best-fit transit model is depicted in Figure \ref{fig:transit_juliet}.

\begin{table}
    \centering
    \begin{tabular}{l c c}
    \hline
    Prior & Description & Distribution \\
    \hline
    \hline
    Period (P) & days & From ExoFOP \\
    $T_0$ & BJD & From ExoFOP \\
    $r_1$ & \citet{2018RNAAS...2..209E} & $\mathcal{U}$(0.0, 1.0)\\
    $r_2$ & \citet{2018RNAAS...2..209E} & $\mathcal{U}$(0.0, 1.0)\\
    a/$R_{\star}$ & & $\mathcal{U}$(1, 100)\\
    Eccentricity & & 0 (Fixed)\\
    $\omega$ & deg & 90 (Fixed)\\
     & & \\
    \multicolumn{3}{c}{Instrumental Parameters}\\
    \hline
     & & \\
    $q_1$ & \citet{2013MNRAS.435.2152K} & $\mathcal{U}$(0.0, 1.0)\\
    $q_2$ & \citet{2013MNRAS.435.2152K} & $\mathcal{U}$(0.0, 1.0)\\
    $m_{flux}$ & ppm & $\mathcal{N}$(0.0, 0.1)\\
    $m_{dilution}$ & & 1.0\\
    $\sigma_{w}$ & ppm & $\mathcal{L}$(0.1, 1000)\\
    \hline
    \end{tabular}
    \caption{Priors provided to \texttt{Juliet} for modeling. We fixed eccentricity to 0 and argument of periastrone to 90$^{^{\circ}}$ for all the planets.}
    $\mathcal{N}$: Normal Distribution\\
    $\mathcal{U}$: Uniform Distribution\\
    $\mathcal{L}$: Log-uniform Distribution\\
    \label{tab:priors}
\end{table}

Relatively dim and/or crowded target stars were often subject to background over-correction in the SPOC pipeline versions employed during the primary TESS mission. This was characterized by overestimated background flux values and, consequently, overestimated transit depths \citep{2020AJ....160..153B}. The SPOC light curves for the first year of the primary mission (sectors 1--13) have been reprocessed with an updated background correction algorithm, but the light curves currently available at MAST for the second year of the primary mission (sectors 14--26) remain susceptible to over-correction of the background level. This could potentially impact our planets TOI-1467b, TOI-1739b, and TOI-2068b. However, for TOI-1739 and TOI-2068, the bias is negligible, being many times less (15$\times$ for TOI-1739 and 50$\times$ for TOI-2068) than the uncertainties in the derived planet radius. Therefore, the only planet affected by the bias is TOI-1467b. Correcting this planet's data involves adding a constant flux value to the \texttt{PDCSAP} light curve for all cadences in sectors 17 and 18. These flux values adjust for the over-correction of the background level in each sector and account for the number of pixels in the photometric aperture. As described in the TESS sector~27 (DR38) data release notes\footnote{\url{https://archive.stsci.edu/missions/tess/doc/tess_drn/tess_sector_27_drn38_v02.pdf}}, adjusted flux values can be calculated using,

\begin{equation}
    \textrm{flux adjustment} = bg_{bias} \times N_{\textrm{optimal aperture}} \times \frac{CROWDSAP}{FLFRCSAP} 
\end{equation}

Here, $bg_{bias}$ is background-bias, $N_{\textrm{optimal aperture}}$ is the number of pixels in the optimal aperture, CROWDSAP and FLFRCSAP are the crowding metric and flux fraction correction reported in the light curve and target pixel files, respectively. These values for sectors 17 and 18 of planet TOI-1467b along with the adjusted flux value are listed in Table \ref{tab:flux_adjust}.

\renewcommand{\arraystretch}{1.2}
\begin{table}
    \centering
    \begin{tabular}{c c c}
         \hline
         Parameters & Sector 17 & Sector 18 \\
         \hline
         \hline
         $bg_{bias}$ ($e^{-} sec^{-1}$) & 12.980  & 21.793 \\
         $N_{\textrm{optimal aperture}}$ (pixels) & 13 & 17 \\
         CROWDSAP & 0.9842 & 0.9643 \\
         FLFRCSAP & 0.8826 & 0.8689 \\
         & & \\
         Flux Adjustment ($e^{-} sec^{-1}$) & 188.16 & 410.68 \\
         \hline
    \end{tabular}
    \caption{Adjusted flux values for TOI-1467.}
    \label{tab:flux_adjust}
\end{table}
\renewcommand{\arraystretch}{1}

We added these adjusted flux values to the \texttt{PDCSAP} flux of sectors 17 and 18. Without flux correction, we derived radius of TOI-1467b to be 1.855$^{+0.112} _{-0.112}$, after flux adjustment it reduced to 1.833$^{+0.159} _{-0.156}$, which represents a 1.17\% reduction. Subsequently, the derived planetary and orbital parameters for each system are represented in Table \ref{tab:params}. We estimated planetary mass using \citet{2017ApJ...834...17C} mass-radius relationship and based on the results we also estimated the planetary density and radial velocity semi-amplitude.

\renewcommand{\arraystretch}{1.5}
\begin{table*}
    \centering
    \resizebox{0.8\textwidth}{!}
    {
    \begin{tabular}{l l c c c c}
    \hline
    Parameters & Units & TOI-238b & TOI-771b & TOI-871b & TOI-1467b \\
    \hline
    \hline
    $P$ & Period (days) & 1.273114$^{+0.000002} _{-0.000002}$ & 2.326021$^{+0.000001} _{-0.000001}$ & 14.362565$^{+0.00009} _{-0.00009}$ & 5.971143$^{+0.000006} _{-0.000006}$ \\
    
    $R_p$ & Radius (Earth Radius) & 1.612$^{+0.096} _{-0.099}$ & 1.422$^{+0.108} _{-0.086}$ & 1.664$^{+0.114} _{-0.113}$ & 1.833$^{+0.159} _{-0.156}$ \\
    
    $T_C$ & Epoch Time (BJD) & 2458354.6608$^{+0.0010} _{-0.0008}$ & 2458572.4178$^{+0.0003} _{-0.0003}$ & 2458417.0523$^{+0.0027} _{-0.0029}$ & 2458766.9895$^{+0.0008} _{-0.0008}$ \\
    
    $T_{dur}$ & Transit Duration (days) & 0.060$^{+0.019} _{-0.015}$ & 0.042$^{+0.019} _{-0.013}$ & 0.168$^{+0.232} _{-0.15}$ & 0.063$^{+0.019} _{-0.015}$ \\
    
    $a$ & Semi-major Axis (AU) & 0.0212$^{+0.0004} _{-0.0004}$ & 0.0207$^{+0.0008} _{-0.0008}$ & 0.1054$^{+0.0021} _{-0.0021}$ & 0.0510$^{+0.0009} _{-0.0009}$ \\
    
    $i$ & Inclination (Degrees) & 85.78$^{+2.21} _{-3.42}$ & 88.13$^{+1.25} _{-1.69}$ & 89.26$^{+0.47} _{-3.86}$ & 88.98$^{+0.44} _{-0.73}$ \\
    
    $T_{eq}$ & Equilibrium Temperature (K) & 1456.61$^{+47.39} _{-47.39}$ & 527.19$^{+22.08} _{-22.08}$ & 621.00$^{+16.05} _{-16.05}$ & 558.89$^{+16.37} _{-16.37}$ \\
    
    $S$ & Insolation ($S_E$) & 747.65$^{+55.61} _{-55.61}$ & 12.86$^{+1.35} _{-1.35}$ & 24.78$^{+1.22} _{-1.22}$ & 16.27$^{+1.21} _{-1.21}$ \\
    
    $R_P/R_{\star}$ & Radius of planet in stellar radii & 0.020$^{+0.001} _{-0.001}$ & 0.054$^{+0.003} _{-0.002}$ & 0.021$^{+0.001} _{-0.001}$ & 0.036$^{+0.002} _{-0.002}$ \\
    
    $a/R_{\star}$ & Semi-major axis in stellar radii & 6.14$^{+0.58} _{-1.04}$ & 15.69$^{+2.18} _{-3.51}$ & 26.30$^{+6.92} _{-14.77}$ & 26.89$^{+2.36} _{-4.46}$ \\
    
    $\delta$ & Transit Depth (Fraction) & 0.00039$^{+0.00004} _{-0.00004}$ & 0.00289$^{+0.00033} _{-0.00020}$ & 0.00045$^{+0.00005} _{-0.00005}$ & 0.00126$^{+0.00012} _{-0.00012}$ \\
    
    $b$ & Impact Parameter & 0.46$^{+0.22} _{-0.22}$ & 0.51$^{+0.25} _{-0.32}$ & 0.30$^{+0.72} _{-0.19}$ & 0.48$^{+0.21} _{-0.19}$ \\
    
    $u_1$ & Limb Darkening coeff & 0.27$^{+0.34} _{-0.18}$ & 0.36$^{+0.31} _{-0.22}$ & 1.04$^{+0.41} _{-0.56}$ & 0.31$^{+0.38} _{-0.23}$ \\
    
    $u_2$ & Limb Darkening coeff & 0.21$^{+0.33} _{-0.31}$ & 0.22$^{+0.32} _{-0.36}$ & -0.24$^{+0.48} _{-0.36}$ & 0.06$^{+0.34} _{-0.25}$ \\ 
    & & & & &\\
    \multicolumn{6}{c}{Estimated Parameters} \\
    $M_p$ & Mass (Earth Mass) & 3.6$^{+2.4} _{-1.3}$ & 2.8$^{+2.0} _{-0.9}$ & 3.8$^{+2.7} _{-1.4}$ & 4.4$^{+3.2} _{-1.7}$ \\
    $\rho$ & Density (cgs) & 4.7$^{+3.3} _{-1.9}$ & 5.4$^{+4.0} _{-2.0}$ & 4.5$^{+3.4} _{-1.9}$ & 3.9$^{+3.0} _{-1.8}$ \\
    $K$ & Radial Velocity Semi-Amplitude (m s$^{-1}$) & 2.5$^{+1.5} _{-0.9}$ & 3.7$^{+2.2} _{-1.0}$ & 1.2$^{+0.8} _{-0.4}$ & 2.5$^{+1.6} _{-0.9}$ \\
    & & & & &\\
    \hline
    & & TOI-1739b & TOI-2068b & TOI-4559b & TOI-5799b \\
    \hline
    \hline
    $P$ &  Period (days) & 8.303342$^{+0.000011} _{-0.000013}$ & 7.768915$^{+0.000025} _{-0.000037}$ & 3.965991$^{+0.000314} _{-0.000332}$ & 4.164753$^{+0.00038} _{-0.000427}$ \\
    
    $R_p$ &  Radius (Earth Radius) & 1.695$^{+0.098} _{-0.085}$ & 1.821$^{+0.162} _{-0.149}$ & 1.415$^{+0.126} _{-0.112}$ & 1.625$^{+0.192} _{-0.128}$ \\
    
    $T_C$ &  Epoch Time (BJD) & 2458685.2395$^{+0.0009} _{-0.0009}$ & 2458683.4258$^{+0.0028} _{-0.0025}$ & 2459335.3447$^{+0.0009} _{-0.0010}$ & 2459772.3341$^{+0.0013} _{-0.0013}$ \\
    
    $T_{dur}$ &  Transit Duration (days) & 0.075$^{+0.024} _{-0.018}$ & 0.086$^{+0.036} _{-0.027}$ & 0.058$^{+0.028} _{-0.019}$ & 0.059$^{+0.031} _{-0.020}$ \\
    
    $a$ &  Semi-major Axis (AU) & 0.0742$^{+0.0014} _{-0.0014}$ & 0.0632$^{+0.0011} _{-0.0011}$ & 0.0359$^{+0.0008} _{-0.0008}$ & 0.0352$^{+0.0011} _{-0.0011}$ \\
    
    $i$ &  Inclination (Degrees) & 89.27$^{+0.47} _{-0.70}$ & 89.01$^{+0.68} _{-1.02}$ & 88.64$^{+0.96} _{-1.49}$ & 88.67$^{+0.89} _{-1.65}$ \\
    
    $T_{eq}$ &  Equilibrium Temperature (K) & 755.55$^{+20.19} _{-20.19}$ & 520.62$^{+14.18} _{-14.18}$ & 554.16$^{+19.14} _{-19.14}$ & 518.01$^{+20.15} _{-20.15}$ \\
    
    $S$ &  Insolation ($S_E$) & 54.35$^{+3.59} _{-3.59}$ & 12.23$^{+0.73} _{-0.73}$ & 15.77$^{+1.44} _{-1.44}$ & 11.91$^{+1.29} _{-1.29}$ \\
    
    $R_P/R_{\star}$ &  Radius of planet in stellar radii & 0.021$^{+0.001} _{-0.001}$ & 0.031$^{+0.002} _{-0.002}$ & 0.035$^{+0.003} _{-0.002}$ & 0.045$^{+0.005} _{-0.003}$ \\
    
    $a/R_{\star}$ &  Semi-major axis in stellar radii & 32.46$^{+2.7} _{-5.87}$ & 26.11$^{+3.9} _{-5.7}$ & 19.71$^{+2.61} _{-4.82}$ & 20.31$^{+2.72} _{-5.43}$ \\
    
    $\delta$ &  Transit Depth (Fraction) & 0.00043$^{+0.00004} _{-0.00003}$ & 0.00097$^{+0.00015} _{-0.00014}$ & 0.0012$^{+0.00018} _{-0.00016}$ & 0.00206$^{+0.00045} _{-0.00027}$ \\
    
    $b$ &  Impact Parameter & 0.41$^{+0.25} _{-0.25}$ & 0.46$^{+0.26} _{-0.30}$ & 0.47$^{+0.28} _{-0.31}$ & 0.48$^{+0.29} _{-0.31}$ \\
    
    $u_1$ &  Limb Darkening coeff & 0.40$^{+0.36} _{-0.27}$ & 0.62$^{+0.50} _{-0.40}$ & 0.40$^{+0.40} _{-0.26}$ & 0.63$^{+0.46} _{-0.40}$ \\
    
    $u_2$ &  Limb Darkening coeff & 0.10$^{+0.35} _{-0.32}$ & -0.07$^{+0.47} _{-0.33}$ & 0.06$^{+0.34} _{-0.30}$ & 0.01$^{+0.42} _{-0.36}$ \\

    & & & & &\\
    
    \multicolumn{6}{c}{Estimated Parameters} \\
    
    $M_p$ & Mass (Earth Mass) & 4.0$^{+2.7} _{-1.5}$ & 4.4$^{+3.2} _{-1.6}$ & 2.7$^{+2.0} _{-1.0}$ & 3.7$^{+2.7} _{-1.4}$ \\
    
    $\rho$ & Density (cgs) & 4.5$^{+3.2} _{-1.8}$ & 4.0$^{+3.1} _{-1.8}$ & 5.3$^{+4.1} _{-2.3}$ & 4.7$^{+3.9} _{-2.1}$ \\
    
    $K$ & Radial Velocity Semi-Amplitude (m s$^{-1}$) & 1.5$^{+0.9} _{-0.5}$ & 2.1$^{+1.4} _{-0.7}$ & 2.0$^{+1.4} _{-0.7}$ & 3.0$^{+2.0} _{-1.0}$ \\
    \hline
    \end{tabular}
    }
    \caption{Median values and 68\% confidence interval for all validated planets from \texttt{Juliet}.}
    \label{tab:params}
\end{table*}

\renewcommand{\arraystretch}{1}

\section{Discussion}\label{sec:discussion}

\subsection{Radius Valley}
The radius valley \citep{2017AJ....154..109F}, reveals a scarcity of planets with sizes between 1.5 and 2.0~R$_{\oplus}$ and orbital periods shorter than 100 days. Most of the planets we have validateded are situated just within this sparsely populated area of parameter space.  Photoevaporation may be a possible cause for the existence of the radius valley \citep{2012ApJ...761...59L,2013ApJ...775..105O,2013ApJ...776....2L,2014A&A...562A..80K,2015AsBio..15...57L,2020A&A...638A..52M}. In a photoevaporation scenario, X-ray and/or extreme ultraviolet radiation from the host star causes the gassy layers of a larger planet to evaporate, leaving behind only a rocky core. 


The position and characteristics of the radius valley have been found to be influenced by the properties of the host star, as demonstrated in previous studies \citep{2017AJ....154..109F,2019MNRAS.487...24G,2020AJ....160..108B}. \citet{2020AJ....159..211C} and \citet{2018MNRAS.479.4786V} revisited this phenomenon in the context of low-mass stars. That study was focused on planets that orbit stars cooler than 4700K and observed that the slope of the radius valley is different from that of FGK stars, and the peaks of the planet size distributions shift towards smaller planets. As a consequence, the center of the radius valley also moved towards smaller planet sizes. Their study considered planets discovered by Kepler and K2 and accounted for any gaps in the data. In essence, this research introduced the concept of 'keystone planets,' which occupy the space between the measured radius valley for low-mass stars and the one previously measured by \citet{2019ApJ...875...29M} for Sun-like stars. These keystone planets play a pivotal role in advancing our understanding of the radius-valley phenomenon around low-mass stars.

\begin{figure*}
    \centering
    \includegraphics[scale = 0.5]{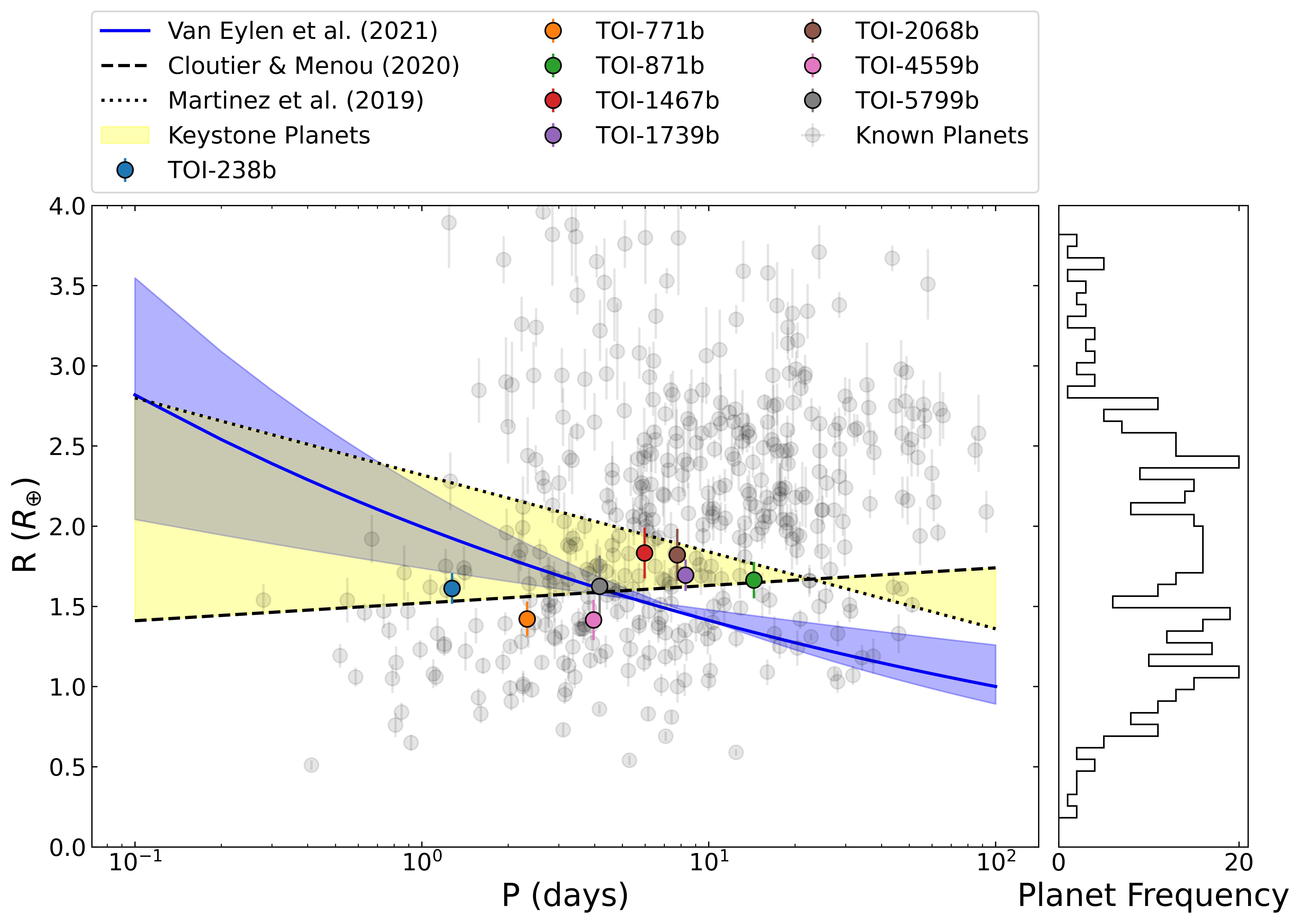}
    \caption{Plot illustrating the currently known sample of planets that orbit stars with temperatures cooler than 5000K in a radius-period space. Gray circles represent confirmed planets with radii measured to better than 10\% precision. The black dashed line corresponds to the low-mass star radius valley for hosts cooler than 4700 K, as determined by \citet{2020AJ....159..211C}, while the dotted black line represents the radius valley for Sun-like stars, as measured by \citet{2019ApJ...875...29M}. The yellow shaded area indicates the region where planets commonly referred to as 'keystone planets' are typically found. The blue line illustrates the radius valley calculated by \citet{2018MNRAS.479.4786V} for hosts cooler than 4000 K.}
    \label{fig:radius_valley}
\end{figure*}

In Figure \ref{fig:radius_valley} we present the current sample of exoplanets orbiting stars cooler than 5000K with planetary radii measured to better than 10\% precision. We show the positions of the radius valleys as measured by \citet{2019ApJ...875...29M}, \citet{2020AJ....159..211C} and \citet{2018MNRAS.479.4786V} and the positions of all eight validated planets. Two of our validated planets, TOI-771b and TOI-4559b, with radii of 1.422~R$_\oplus$ and 1.415~R${_\oplus}$ respectively, fall just below the 1.56~R$_\oplus$ boundary for a 2.326~day orbital period and the 1.58~R$_\oplus$ boundary for a 3.966~day orbital period. Hence, both of these planets lie within the region of likely rocky planets. Hence they can be considered as super-Earths. Our other six planets (i.e., TOI-238b, TOI-871b, TOI-1467b, TOI-1739b, TOI-2068b and TOI-5799b) fall within the keystone region. The thermally driven atmospheric mass loss predicts that planets within the keystone region should be predominantly rocky, conversely the gas-poor formation scenario (whose radius valley slope differs in sign from that of thermally driven mass loss) predicted that those planets should be primarily non-rocky. To validate these hypotheses effectively, a rigorous approach involves the selection of planets located within the region of interest and the acquisition of precise measurements related to their bulk density \citep{2020AJ....159..211C}. Thus, future mass measurements for these planets will enable us to refine our understanding of the radius valley phenomenon around low-mass stars, further elucidate the connections between planet size and host star properties, and contribute valuable insights into the formation and evolution of planetary systems in different stellar environments.

\subsection{Prospect for Radial Velocity Follow-up}\label{sec:RV Follow-up}
Masses are important to determine the composition of small planets as well as to better interpret their atmospheric compositions and formation mechanisms. Employing high-precision measurements of radial velocity (RV) would not solely serve to restrict the planetary mass, but also serve to delimit its orbital parameters, including eccentricity, thereby affording insights into the system's orbital dynamics. We derived the possible RV semi-amplitudes for all the validated planets using the mass estimated via mass-radius relationship mentioned in \citet{2017ApJ...834...17C}. These values are listed in Table \ref{tab:params}.

In order to detect RV signals for our validated planets, a precision of at least 1 m s$^{-1}$ or finer is required. Achieving such a high level of precision can be quite challenging for many instruments. However, for those planets visible from the southern hemisphere, such as TOI-238b, TOI-771b, TOI-871b, and TOI-4559, we have access to two spectrographs: ESPRESSO at the Very Large Telescope \citep[VLT;][]{2013Msngr.153....6P}, located at the Paranal Observatory, and HARPS \citep{2000SPIE.4008..582P} at the La Silla Observatory. ESPRESSO stands out with its ability to detect long term ($\sim$1 year) RV signals with a precision as fine as 0.5 m s$^{-1}$, provided the visual magnitude 14 mag or brighter. This spectrograph is ideally suited for observing non-active, non-rotating, and quiet G dwarfs to red dwarfs, making it an excellent choice for such candidates. HARPS, on the other hand, is also a formidable instrument for capturing subtle RV signals, offering a precision of 1 m s$^{-1}$.

For planets situated in the northern hemisphere, such as TOI-1467b, TOI-1739b, TOI-2068, and TOI-5799b, there are a number of instruments that are uniquely suited to radial velocity mass measurements: CARMENES at the Calar Alto Observatory \citep{2020SPIE11447E..3CQ}, NEID at the Kitt Peak National Observatory \citep{2016SPIE.9908E..7HS}, EXPRES at the Lowell Observatory \citep{2016SPIE.9908E..6TJ}, and MAROON-X at the Gemini-N \citep{2018SPIE10702E..6DS}. CARMENES can detect RV signals with a precision of 1 m s$^{-1}$ when the S/N is 150, assuming the magnitude limit does not exceed 10.5 on the J-band. Given the J magnitudes of these northern hemisphere planets in Table \ref{tab:stellar}, all of them might be observable using CARMENES. NEID, another excellent instrument, has the capability to observe with a precision of 1 m s$^{-1}$. EXPRES offers even finer precision, allowing observations at 0.3~m~s$^{-1}$ with a remarkable S/N of 250~pixel$^{-1}$. MAROON-X, located at the Gemini-N, excels in capturing signals with a precision of less than 1~m~s$^{-1}$, particularly well-suited for M-dwarfs, provided that the visual magnitude remains below 16 mag.

MAROON-X is particularly suited to obtain high-precision RVs for M dwarf hosts due to its broad red-optical wavelength coverage. Assuming an exposure time of 1800~s and excellent weather conditions, the predicted RV precision achievable on targets are listed in Table \ref{tab:maroon-x} for blue and red arms. Based on the results we obtained it could be possible to measure the masses of the three transiting planets TOI-238b, TOI-1467b and TOI-1739 using MAROON-X. The number of observations needed depends on how precise the mass determination to be, the actual RV precision achieved, and the amount of stellar variability in the spectroscopic data.

\begin{table}
    \centering
    \begin{tabular}{c c c}
    \hline
    Planet & \multicolumn{2}{c}{RV Precision (m s$^{-1}$)} \\
           & Blue Arm  & Red Arm \\
    \hline
    \hline
    TOI-238b & 0.6 & 0.9 \\
    TOI-1467b & 1.9 & 1.8 \\
    TOI-1739b & 0.5 & 0.9 \\
    TOI-2068b & 2.9 & 2.6 \\
    TOI-5799b & 3.6 & 3.0 \\
    \hline
    \end{tabular}
    \caption{Photon noise limited RV uncertainties for detectable targets using MAROON-X at Gemini-N.}
    \label{tab:maroon-x}
\end{table}

It is worth noting that the estimations for the RVs described in Table \ref{tab:params} refer to the case of circular orbits, in case of eccentric orbit the induced RVs semi-amplitudes would be slightly larger and hence easier to detect.

\subsection{Transmission and Emission Spectroscopy}
\citet{2018PASP..130k4401K} introduced a methodology for computing the Transmission and Emission Spectroscopy Metrics (TSM and ESM). These metrics are determined by considering the luminosity of the host star, the planetary radius, mass, and equilibrium temperature. By evaluating the anticipated signal-to-noise ratio of observations with the James Webb Space Telescope \citep[JWST;][]{2006SSRv..123..485G} for both transmission and emission spectroscopy of a given exoplanet, TSM and ESM serve as valuable tools for identifying the most promising targets for atmospheric characterization among the exoplanets discovered by the Transiting Exoplanet Survey Satellite (TESS).

\begin{figure*}
    \centering
    \includegraphics[scale = 0.35]{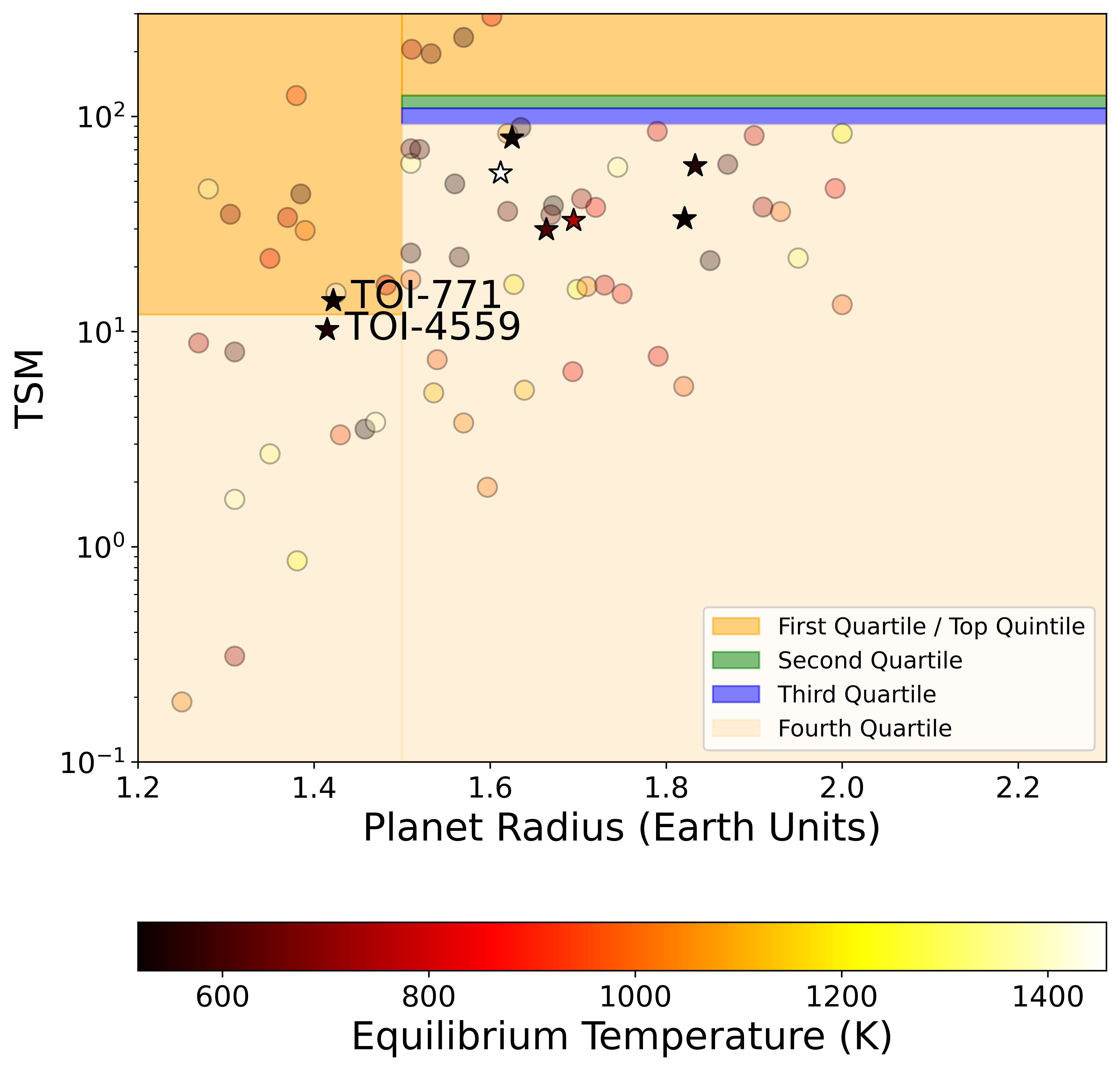}
    \includegraphics[scale = 0.35]{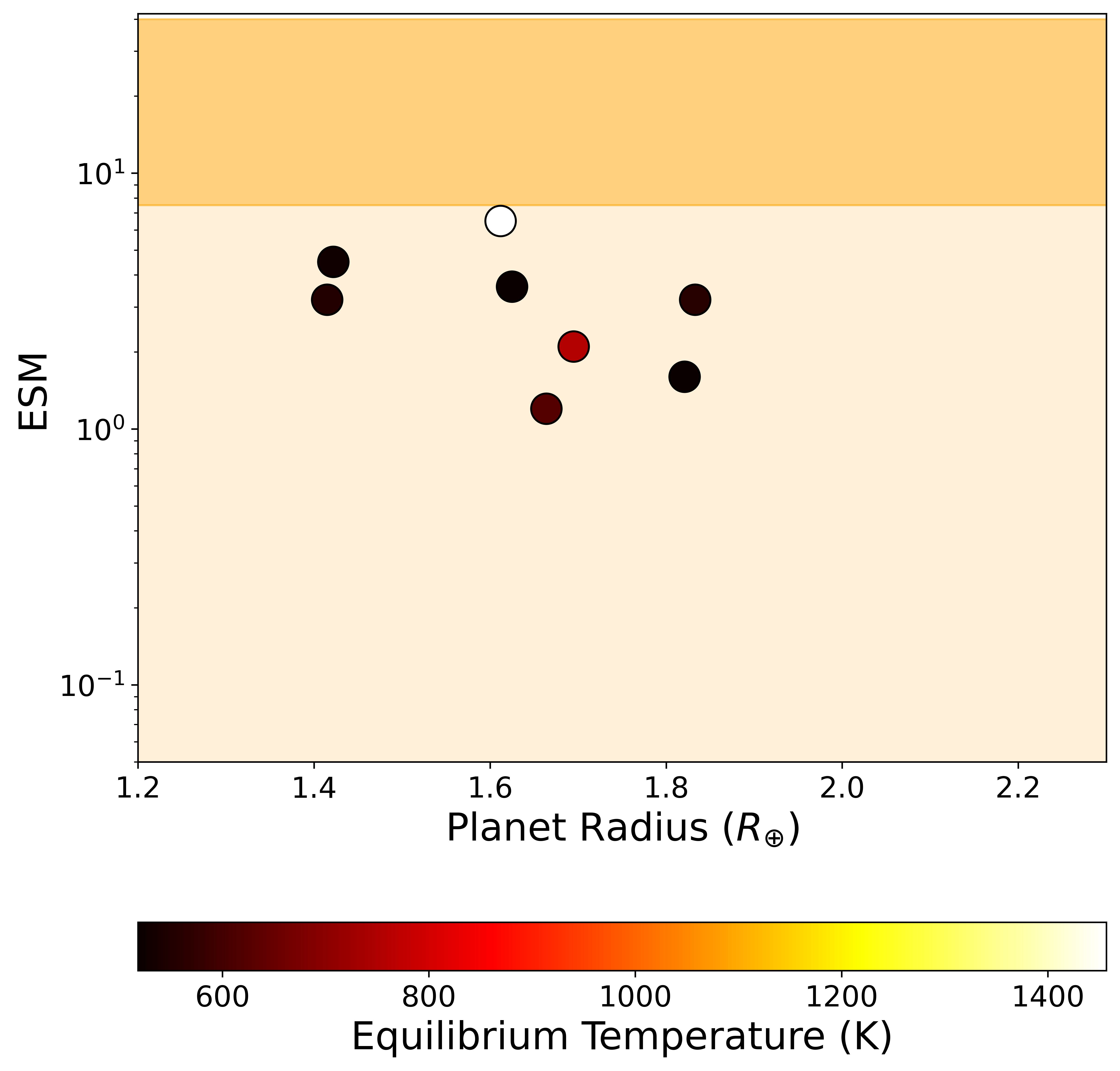}
    \caption{Transmission (Left) and Emission (Right) spectroscopy metrics plotted against the planetary radii. Planets with labels are amenable for transmission or emission spectroscopy using JWST.}
    \label{fig:TSM_ESM}
\end{figure*}

As can be seen from Figure \ref{fig:TSM_ESM}, we find two super-Earth (TOI-771b and TOI-4559b) from our study that are above the threshold of the first quartile suggested by \citet{2018PASP..130k4401K} that makes it a good target for transmission spectroscopy using JWST. On the other hand, the other planets are not amenable either to transmission or emission spectroscopy. We list them in Table \ref{tab:TSM_ESM_known}. Such studies could directly test our hypotheses about the planet's bulk composition and formation history by assessing the elemental compositions and total metal enrichment of the planet's atmosphere. Future mass measurements and spectral analyses will be instrumental in ascertaining the atmospheric composition of this particular super-Earth, as well as others like it.

\begin{table}
    \centering{
    \begin{tabular}{c c c c c}
        \hline
        Planet & TSM & ESM & TS\footnotemark[1] & ES\footnotemark[2] \\
        \hline
        \hline
        TOI-238b & 54.5 & 6.5 & & \\
        TOI-771b & 13.9 & 4.5 & Y & \\
        TOI-871b & 29.7 & 1.2 & & \\
        TOI-1467b & 58.9 & 3.2 & & \\
        TOI-1739b & 32.8 & 2.1 & & \\
        TOI-2068b & 33.4 & 1.6 & & \\
        TOI-4559b & 10.2 & 3.2 & Y & \\
        TOI-5799b & 79.2 & 3.6 & & \\
        \hline
    \end{tabular}} \\
    \footnotemark[1]{Planets amenable for transmission spectroscopy with JWST.}
    \footnotemark[2]{Planets amenable for emission spectroscopy with JWST.}
    \caption{Transmission and Emission spectroscopy metrics for planets validated in this study.}
    \label{tab:TSM_ESM_known}
\end{table}

\subsection{Cosmic Shoreline}
Integrated extreme ultraviolet (XUV) stellar radiation intercepted by a planet (also known as Insolation, $I_{XUV}$) and surface gravity of a planet follow a $I_{XUV} \propto v_{esc}^{4}$ power law \citep{2017ApJ...843..122Z}. The boundary line is called the ``cosmic shoreline". The cosmic shoreline is a hypothesis that suggests that there should be some relation between planetary mass and XUV irradiation that defines a boundary between planets with and without an atmosphere. The shoreline hypothesis suggests investigating the extent to which the escape process influences how planets manage their volatile materials In other words, where the gravitational pull is stronger (high $v_{esc}$) or stellar influence is weaker (low $I_{XUV}$), planets tend to have thick atmospheres. On the other hand, planets less likely to hold an atmosphere are found where gravity is weak or the star is too bright.

\begin{figure*}
    \centering
    \includegraphics[scale = 0.4]{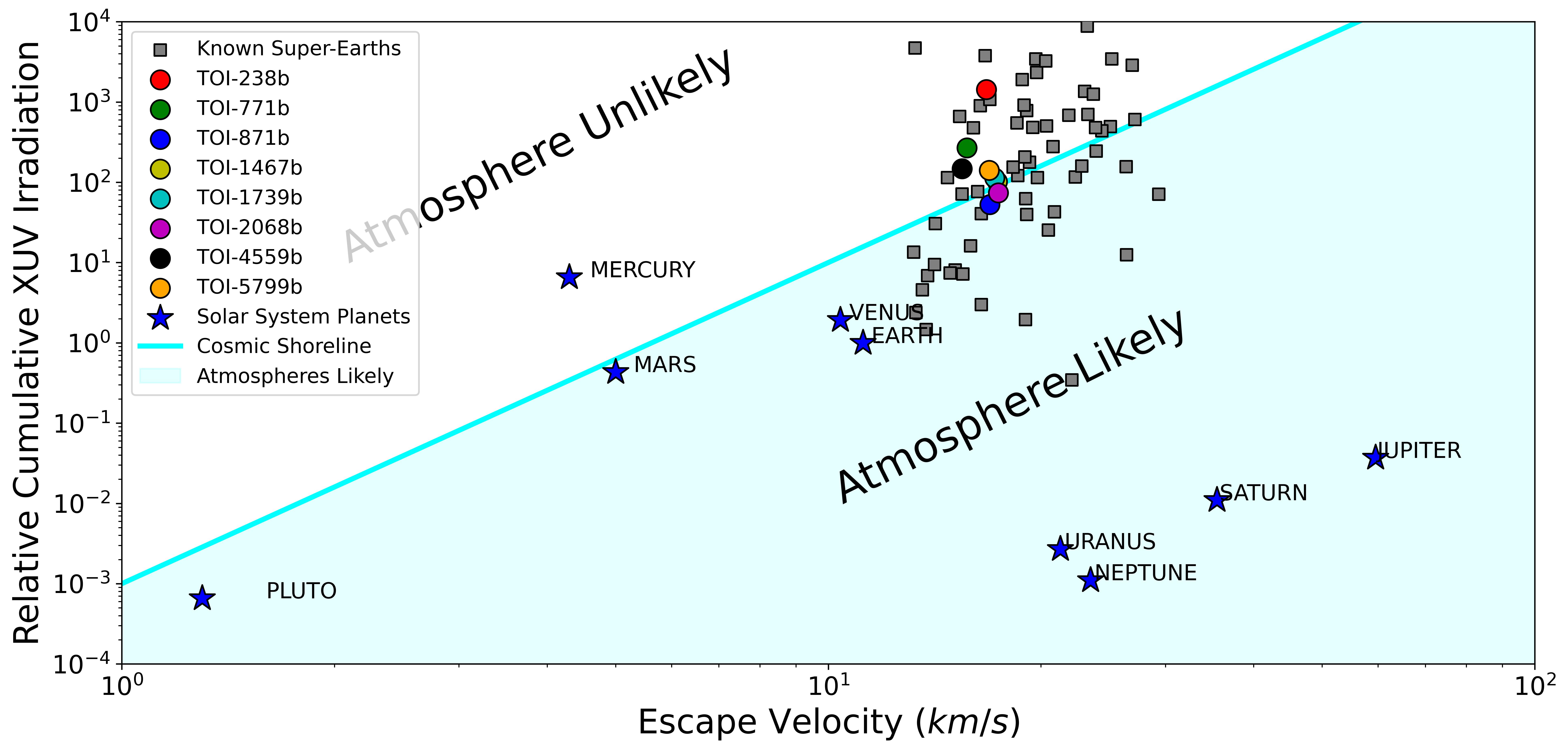}
    \caption{Cumulative XUV irradiation, which is thought to be the driving force behind planetary evaporation vs escape velocity to represent the magnitude of gravity. The atmosphere can be found where gravity is high and solar irradiation is low. The light blue colored line represents the ``cosmic shoreline'' which follows a power law, $I_{XUV} \propto v_{esc}^{4}$ \citep{2017ApJ...843..122Z}. The shaded region (on the right side of the shoreline) contains the planets thought to likely have atmospheres, while the other side represents the planets thought to be unlikely to possess atmospheres. Square boxes represent the known super-Earths.}
    \label{fig:cosmic_shoreline}
\end{figure*}

Figure \ref{fig:cosmic_shoreline} represents the scatter of solar system planets (star marker), known super-Earths (square markers), and our eight validated planets (circle markers). The light blue diagonal line represents the cosmic shoreline. From the eight validated planets, three of them (TOI-871b, TOI-1467b, and TOI-2068b) lie at the right side of a cosmic shoreline, which suggests that they likely hold an atmosphere. And they almost share the same place in the plot thus providing an excellent opportunity to explore the atmospheres of small planets that evolved in the similar stellar environment. If we discover that any of the planets hold an atmosphere, we can set an upper limit on the location of the shoreline. On the other hand, if the planets don't align with the hypothesis, it would indicate that other processes play a more significant role.

\section{Conclusion}\label{sec:conclusion}
In this study, we validate eight exoplanets using TESS, ground-based transit photometry, high-resolution imaging, and a statistical validation tool. Two of our validated planets, TOI-771b and TOI-4559b, have a radius of 1.422~R$_\oplus$ and 1.415~R$_{\oplus}$, respectively. These measurements place them just below the 1.56 R$_\oplus$ radius valley boundary for a 2.326~day orbital period and the 1.58 R$_\oplus$ boundary for a 3.966~day orbital period. Hence, both of these planets align well with the likely rocky composition. However, future mass measurements will provide better constraints on the composition and physical properties of these planets. TOI-238b (1456 K) is one of the hottest exoplanets discovered by TESS, followed by TOI-451b \citep[1491 K;][]{2021AJ....161...65N}, TOI-1416b \citep[1517 K;][]{2023A&A...677A..12D}, TOI-1860b \citep[1885 K;][]{2022AJ....163...99G}, HD 93963 Ab \citep[2042 K;][]{2022A&A...667A...1S}, HD 213885b \citep[2128 K;][]{2020MNRAS.491.2982E}, HD 20329b \citep[2141 K;][]{2022A&A...668A.158M}, TOI-561b \citep[2310 K;][]{2022MNRAS.511.4551L} and TOI-2260b \citep[2609 K;][]{2022AJ....163...99G}. We found that, though challenging, it could be possible to measure the masses of the three transiting planets TOI-238b, TOI-1467b, and TOI-1739 using MAROON-X, located at Gemini North. We also found that two of our validated planets, TOI-771b, and TOI-4559b, are amenable for transmission spectroscopy using JWST.

\begin{acknowledgement}
This work makes use of observations from the LCOGT network. Part of the LCOGT telescope time was granted by NOIRLab through the Mid-Scale Innovations Program (MSIP). MSIP is funded by NSF. This research has made use of the Exoplanet Follow-up Observation Program (ExoFOP; DOI: 10.26134/ExoFOP5) website, which is operated by the California Institute of Technology, under contract with the National Aeronautics and Space Administration under the Exoplanet Exploration Program. Funding for the TESS mission is provided by NASA's Science Mission Directorate. KAC and CNW acknowledge support from the TESS mission via subaward s3449 from MIT. This paper is based on observations made with the MuSCAT3 instrument, developed by the Astrobiology Center and under financial support by JSPS KAKENHI (JP18H05439) and JST PRESTO (JPMJPR1775), at Faulkes Telescope North on Maui, HI, operated by the Las Cumbres Observatory. This paper makes use of observations made with the MuSCAT2 instrument, developed by the Astrobiology Center, at TCS operated on the island of Tenerife by the IAC in the Spanish Observatorio del Teide. This paper makes use of data from the MEarth Project, which is a collaboration between Harvard University and the Smithsonian Astrophysical Observatory. The MEarth Project acknowledges funding from the David and Lucile Packard Fellowship for Science and Engineering, the National Science Foundation under grants AST-0807690, AST-1109468, AST-1616624 and AST-1004488 (Alan T. Waterman Award), the National Aeronautics and Space Administration under Grant No. 80NSSC18K0476 issued through the XRP Program, and the John Templeton Foundation. This work has made use of data from the European Space Agency (ESA) mission {\it Gaia} (\url{https://www.cosmos.esa.int/gaia}), processed by the {\it Gaia} Data Processing and Analysis Consortium (DPAC, \url{https://www.cosmos.esa.int/web/gaia/dpac/consortium}). Funding for the DPAC has been provided by national institutions, in particular, the institutions participating in the {\it Gaia} Multilateral Agreement. The research leading to these results has received funding from  the ARC grant for Concerted Research Actions, financed by the Wallonia-Brussels Federation. TRAPPIST is funded by the Belgian Fund for Scientific Research (Fond National de la Recherche Scientifique, FNRS) under the grant PDR T.0120.21. TRAPPIST-North is a project funded by the University of Liege (Belgium), in collaboration with Cadi Ayyad University of Marrakech (Morocco). MG is F.R.S.-FNRS Research Director and EJ is F.R.S.-FNRS Senior Research Associate. The postdoctoral fellowship of KB is funded by F.R.S.-FNRS grant T.0109.20 and by the Francqui Foundation. Based on data collected by the SPECULOOS-South Observatory at the ESO Paranal Observatory in Chile.The ULiege's contribution to SPECULOOS has received funding from the European Research Council under the European Union's Seventh Framework Programme (FP/2007-2013) (grant Agreement n$^\circ$ 336480/SPECULOOS), from the Balzan Prize and Francqui Foundations, from the Belgian Scientific Research Foundation (F.R.S.-FNRS; grant n$^\circ$ T.0109.20), from the University of Liege, and from the ARC grant for Concerted Research Actions financed by the Wallonia-Brussels Federation. This work is supported by a grant from the Simons Foundation (PI Queloz, grant number 327127). This research is in part funded by the European Union's Horizon 2020 research and innovation programme (grants agreements n$^{\circ}$ 803193/BEBOP), and from the Science and Technology Facilities Council (STFC; grant n$^\circ$ ST/S00193X/1, and ST/W000385/1). The material is based upon work supported by NASA under award number 80GSFC21M0002. This publication benefits from the support of the French Community of Belgium in the context of the FRIA Doctoral Grant awarded to MT. E. D acknowledges support from the innovation and research Horizon 2020 program in the context of the Marie Sklodowska-Curie subvention 945298. This research has made use of the NASA Exoplanet Archive, which is operated by the California Institute of Technology, under contract with the National Aeronautics and Space Administration under the Exoplanet Exploration Program. We acknowledge the use of public TESS data from pipelines at the TESS Science Office and at the TESS Science Processing Operations Center. Resources supporting this work were provided by the NASA High-End Computing (HEC) Program through the NASA Advanced Supercomputing (NAS) Division at Ames Research Center for the production of the SPOC data products. This research was carried out in part at the Jet Propulsion Laboratory, California Institute of Technology, under a contract with the National Aeronautics and Space Administration (80NM0018D0004). This research was supported by NASA Grant 18-2XRP18\_2-0007 awarded to DRC.
\end{acknowledgement}

\paragraph{Data Availability Statement}
The TESS photometry and the high angular resolution imaging data used in this article are available at the Mikulski Archive for Space Telescope (MAST) (\url{https://mast.stsci.edu/portal/Mashup/Clients/Mast/Portal.html}) and the ExoFOP-TESS website (\url{https://exofop.ipac.caltech.edu/tess/target.php?id=76923707}) respectively. The data underlying this article will be made available on the reasonable request to the corresponding author.

\printendnotes
\printbibliography
\end{document}